\documentclass[a4paper,11pt]{article}
\usepackage{jheppub}

\usepackage[utf8]{inputenc}
\usepackage{bbm}
\usepackage{cancel}
\usepackage{subcaption}
\usepackage{comment}
\usepackage{slashed}
\usepackage{physics}
\usepackage{tikz-cd}
\usepackage{pgfplots}
\usepackage{pgfplotstable}

\newcommand\beq{\begin{equation}}
\newcommand\eeq{\end{equation}}

\newcommand\Dm{\Delta_{\min}}

\begin{document}

\title{Towers of Operators in CFTs and Convexity Bounds at Large Charge}

\author[1]{Fedor K. Popov}
\author[2]{and Adar Sharon}

\affiliation[1]{Simons Center for Geometry and Physics, Stony Brook University, Stony Brook, NY}
\affiliation[2]{Mani L. Bhaumik Institute for Theoretical Physics, Department of Physics and Astronomy,
University of California, Los Angeles, CA 90095, USA}

\emailAdd{fpopov@scgp.stonybrook.edu}
\emailAdd{asharon@physics.ucla.edu}

\abstract{In \cite{Cuomo:2024fuy}, it was shown that for 3d CFTs with a moduli space along which a $U(1)$ symmetry is spontaneously broken, the minimum scaling dimension at large charge $Q$ scales as $\Delta_{\min}(Q)=\alpha_1 Q+\alpha_0+O(1/Q)$. Motivated by the holographic swampland program, we study possible bounds on the coefficients $\alpha_i$. For $\alpha_0$, the weak gravity conjecture motivates the CFT charge convexity conjecture, which requires the bound $\alpha_0\leq 0$. Using the moduli space EFT we prove this bound for the \textit{projected} $\Dm(Q)$, obtained by fixing a single charge $Q$ and minimizing the dimension while allowing all other charges to vary. This provides a WGC-motivated bound that is explicitly provable using CFT methods.
On the other hand, we show that $\alpha_1$ admits no universal bound apart from the trivial bound $\alpha_1\geq 0$. We also compute $\alpha_1$ and $\alpha_0$ in several new 3d $\mathcal{N}=1$ theories via the $\epsilon$-expansion and large-$N$ methods.}

\setcounter{tocdepth}{2}
\maketitle

\newpage

\section{Introduction}

The study of Conformal Field Theories (CFTs) ultimately leads to the study of the allowed spectrum of local operators. One of the central problems in CFTs is to constrain this spectrum using conformal symmetry, locality and unitarity. An especially interesting case is when a subclass of operators arranges itself into a ``tower''. For example, crossing symmetry implies the existence of double-twist towers whose dimensions become asymptotically additive \cite{Komargodski:2012ek,Fitzpatrick:2012yx,Caron-Huot:2017vep}, while infinite towers of conserved higher-spin currents force the theory to be free, or close to free, in a precise sense \cite{Maldacena:2011jn,Maldacena:2012sf}. In holographic CFTs, towers of single-trace operators are the boundary imprint of bulk particle spectra, including Kaluza--Klein modes and string excitations. Thus the structure of large-charge towers gives a CFT way of probing when an extra geometric direction, or a higher-dimensional bulk description, is emerging. Towers also play a central role in the swampland program. The distance conjecture predicts an infinite tower of light states at infinite-distance limits in quantum gravity \cite{Vafa:2005ui,Ooguri:2006in} (see \cite{Palti:2019pca} for a review), and in holographic settings this can be translated into statements about towers of CFT operators \cite{Perlmutter:2020buo,Baume:2023msm,Baume:2020dqd}. Thus, in several different contexts, the existence and properties of towers of operators provide sharp information about the structure of the underlying theory.

The best-known linear towers of non-spinning operators arise in supersymmetric CFTs with four supercharges. In these theories, moduli spaces are associated with towers of chiral operators. If $\Phi$ is a chiral operator of R-charge $r$ parameterizing the moduli space, then $\Phi^n$ is also chiral and has dimension
\begin{equation}\label{eq:BPS}
  \Delta_{\Phi^n}=\frac{d-1}{2}rn.
\end{equation}
However, linear towers are not restricted to this highly supersymmetric setting, and they also arise in theories with less supersymmetry. Recently, \cite{Cuomo:2024fuy} studied general CFTs with moduli spaces along which a continuous global symmetry is spontaneously broken. The main result is that if $Q$ is the charge under a broken $U(1)$, then 
\begin{equation}\label{eq:large-charge-Dm}
  \Dm(Q)=\alpha_1 Q+\cdots,
\end{equation}
where the ellipsis denotes terms which are subleading at large $Q$. Here and below, $\Dm(Q)$ denotes the lowest scaling dimension in the charge-$Q$ sector. In cases where the relevant symmetry is an $R$-symmetry, this result reduces to \eqref{eq:BPS}.

We thus find that moduli spaces with less than four supercharges obey the linear relation \eqref{eq:BPS} only asymptotically at large $Q$. The form of the corrections can also be fixed. Using large-charge methods \cite{Hellerman:2015nra} applied to the moduli-space EFT \cite{Hellerman:2017sur,Hellerman:2017veg,Cuomo:2024fuy}, one finds that in 3d, which will be our main focus, the expansion takes the form
\begin{equation}\label{eq:intro-large-charge-3d}
  \Dm(Q)=\alpha_1 Q+\alpha_0+O(1/Q),
\end{equation}
while in 4d the same moduli-space EFT analysis gives
\begin{equation}\label{eq:intro-large-charge-4d}
  \Dm(Q)=\alpha_1 Q+\alpha_0\log Q+O(Q^0).
\end{equation}
The main examples in this paper are 3d $\mathcal N=1$ theories, which can have moduli spaces despite having only two supercharges, as explained e.g.~in \cite{Gaiotto:2018yjh}. In terms of the EFT analysis we will be using, $\alpha_1$ is a classical contribution while $\alpha_0$ comes from the one-loop correction. These are the two main observables we will be studying in this paper.

There are two related reasons to study these coefficients. The first is simply that there are very few examples in which they can be computed in theories with less than four supercharges. In the first part of this paper we compute $\alpha_1,\alpha_0$ in several new examples using large $N$ or small $\epsilon$ expansions. Some of these computations are the first direct computations of the subleading coefficient in 3d, rather than computations obtained by continuing from the $\epsilon$-expansion. The examples include large-$N_f$ computations in 3d $\mathcal N=1$ SQED, $SU(2)$ SQCD, and Wess-Zumino models. The details of these computations are reviewed in section \ref{sec:computations}.

The second motivation comes from the holographic swampland program, and more specifically from convexity conjectures for charged operators. The idea is to translate expectations from the swampland program into constraints on CFT data by using the AdS/CFT correspondence. Motivated by the weak gravity conjecture (WGC) \cite{ArkaniHamed:2006dz}, Aharony and Palti \cite{Aharony:2021mpc} introduced the CFT convexity conjecture (CCC), which requires $\Dm(Q)$ to be superadditive,
\begin{equation}\label{eq:intro-superadditivity}
  \Dm(Q_1+Q_2)\geq \Dm(Q_1)+\Dm(Q_2),
\end{equation}
for charges which are multiples of some order-one unit charge $q_0$. This conjecture has been tested and refined in several examples and generalizations \cite{Antipin:2021rsh,Aharony:2023cbz,Palti:2022unw}. A large-charge EFT discussion of convexity was given in \cite{Orlando:2023ljh}, while \cite{Cohen:2025slc} found that bottom-up EFT consistency alone is not enough to prove it. Ultimately it was found that this strong version of the conjecture is not correct in general \cite{Sharon:2023wgr}. 

Nevertheless, one can ask for a weaker version of the same statement, where the order-one charge assumption on $q_0$ is dropped and the inequality is required only at parametrically large charge. All known large-charge phases are manifestly consistent with convexity,\footnote{It is certainly possible that there are undiscovered large-charge phases with an exotic asymptotic behavior of $\Dm$, for which convexity would have to be studied separately.} the only exception being the asymptotically linear behavior of \eqref{eq:intro-large-charge-3d} for a 3d $\mathcal{N}=1$ theory with a moduli space. In this case the asymptotic version of \eqref{eq:intro-superadditivity} (known as the asymptotic CCC) requires\footnote{We emphasize that the bound is a necessary but not sufficient condition for the asymptotic CCC, since if $\alpha_0=0$ then the next subleading term in the large-$Q$ expansion \eqref{eq:intro-large-charge-3d} determines convexity. The only known cases with $\alpha_0=0$ have an exactly linear tower (so that all other subleading corrections also vanish and convexity is obeyed), but it is certainly possible that other cases exist.}
\begin{equation}\label{eq:bound_alpha0}
  \text{Asymptotic CCC bound:}\qquad\alpha_0\leq0.
\end{equation}
This is an example of an explicit CFT constraint motivated by holographic swampland, and one can now attempt to prove or disprove it using CFT tools. 

In this paper we prove a version of the bound \eqref{eq:bound_alpha0}. In the case of a single $U(1)$ symmetry our version is equivalent to the desired bound \eqref{eq:bound_alpha0}, with a subtlety arising only when considering symmetry groups with several Cartans. For a set of Cartans $\{Q_a\}_{a=1}^N$, let $\Delta_{\min}(Q_a)$ denote the lowest dimension at fixed charge vector $Q_a$. The stronger fixed-charge-ray convexity statement is that, for every fixed ray in charge space, $\Delta_{\min}(Q_a)$ is asymptotically convex. Although all known CFT examples (including the ones discussed in this paper) are consistent with this statement, we were not able to prove it using large-charge methods. In fact, we will find a 3d $\mathcal{N}=1$ EFT counterexample to this statement (which does not correspond to any known complete CFT). Thus we find that large-charge methods are insufficient to prove this statement in full generality without some additional input from the full UV CFT, in agreement with \cite{Cohen:2025slc}.

Instead, we focus on a slightly weaker statement, which involves a projection to a single Cartan. Given a set of Cartan generators $T^a$ and some specific linear combination of them
\begin{equation}
  T_n=n_aT^a,
\end{equation}
define
\begin{equation}\label{eq:projected_delta}
  \Delta_{\min}^{(n)}(Q)= \min_{n\cdot Q=Q}\Delta_{\min}(Q_a).
\end{equation}
Thus only the charge under $T_n$ is held fixed, while the charges under all orthogonal Cartans are minimized over. In particular, the state which minimizes \eqref{eq:projected_delta} may carry nonzero charges under the Cartans orthogonal to $T_n$.

Our main general result is a proof of the projected convexity bound. In 3d,
\begin{equation}
  \Delta_{\min}^{(n)}(Q) = \alpha_1^{(n)}Q+\alpha_0^{(n)}+O(1/Q),
\end{equation}
and we show that the full one-loop coefficient obeys 
\begin{equation}\label{eq:intro_bound}
  \alpha_0^{(n)}\leq 0
\end{equation} 
for any choice of Cartan generator $T_n$. If the inequality is strict, this implies asymptotic superadditivity of the projected tower. If $\alpha_0^{(n)}=0$, the sign of the next term in the large-charge expansion must be examined. This provides a concrete example of a CFT bound motivated by the holographic swampland program that can be derived directly using CFT tools. We will focus on showing that the scalar contribution to $\alpha_0^{(n)}$ is non-positive, since the fermion and gauge field contributions are always non-positive using an argument from \cite{Cuomo:2024fuy}.

As discussed above, this result is weaker than fixed-charge-ray convexity. The projected observable allows the theory to choose the most favorable charge vector with fixed $n\cdot Q$, while the fixed-charge-ray problem prescribes the full charge direction in advance. In some simple classes of EFTs, including very low-dimensional moduli spaces and maximally symmetric sigma-model targets, one can still prove the fixed-charge-ray sign bound by additional arguments. We discuss these cases in section~\ref{subsec:fixed-ray-limitations}. 

This projected statement is also the natural version to compare with the WGC when there is more than one Cartan. For several $U(1)$'s, the relevant flat-space statement is not a separate one-dimensional WGC along every fixed charge ray, but rather the convex-hull WGC \cite{Cheung:2014vva}: for every direction in the dual charge space, the spectrum must contain states whose charge-to-mass vectors have sufficiently large projection in that direction. In the black-hole discharge argument, convex combinations of such vectors are then interpreted as multi-particle final states. Our projected observable is analogous to the first, lower-envelope part of this statement: it fixes a direction $T_n$ and lets the theory choose the lightest tower among all states with fixed $n\cdot Q$.

Let us also mention the coefficient $\alpha_1$. In a simple holographic picture, the linear tower resembles a Kaluza--Klein tower, with the charge interpreted as momentum along an internal circle. This suggests asking whether $\alpha_1$ is bounded by some analogue of the absence of parametric scale separation.\footnote{We thank Irene Valenzuela for this observation.} Here ``absence of scale separation'' refers to the swampland expectation that controlled AdS quantum gravity vacua with compact extra dimensions cannot have a parametrically large hierarchy between the AdS scale and the lightest Kaluza--Klein scale: $m_{\rm KK}L_{\rm AdS}$ should remain order one. In this sense, the relevant swampland conjectures forbid parametrically scale-separated AdS compactifications
\cite{Gautason:2015tig,Lust:2019zwm,Cribiori:2022nke}; see \cite{COUDARCHET20241} for a recent review. Scale-separated AdS flux vacua have a long history, starting with constructions such as DGKT and related type-IIA orientifold models \cite{DeWolfe:2005uu,Camara:2005dc}, while further discussions of scale separation in type-II AdS flux vacua and holography include
\cite{Font:2019uva,Polchinski:2009ch,deAlwis:2014wia,Conlon:2018eyr,Conlon:2020wmc,Collins:2022nux}. We will argue that there is no such universal bound on $\alpha_1$ itself: simple operations on CFTs can make it parametrically large or small. A more natural normalized quantity, involving current and stress-tensor two-point function coefficients, may still be worth considering, but we will not find a universal bound on $\alpha_1$ alone.

The rest of the paper is organized as follows. In section \ref{sec:computations} we compute $\alpha_1,\alpha_0$ in several double-scaling limits, including large-$N_f$ and $\epsilon$-expansion examples. In section \ref{sec:alpha0} we consider $\alpha_0$. We review the quadratic fluctuation problem of \cite{Cuomo:2024fuy}, prove the projected convexity bound for the one-charge observable, and explain why the argument does not extend to arbitrary fixed charge rays. We then give the EFT counterexample to fixed-charge-ray convexity. In section \ref{sec:alpha1} we discuss possible bounds on $\alpha_1$, and explain why $\alpha_1$ itself is not expected to obey a universal bound.

\section{Computations in double-scaling limits}\label{sec:computations}

In this section we compute the large-charge coefficients $\alpha_1$ and $\alpha_0$ in several controlled double-scaling limits of 3d $\mathcal N=1$ theories. We use large-charge methods applied to the moduli space EFT: by the state-operator correspondence, we work on the cylinder, expand around the homogeneous large-charge saddle, and compute its energy. Throughout this paper we set the cylinder radius to one. The leading coefficient $\alpha_1$ comes from the classical saddle energy, while $\alpha_0$ comes from the one-loop Casimir energy of fluctuations. The examples include large-$N_f$ limits and $\epsilon$-expansions. Our superspace conventions follow \cite{Gates:1983nr}, while the moduli-space EFT and large-charge normalization follow \cite{Cuomo:2024fuy}. The argument for the protection of certain moduli spaces of 3d $\mathcal{N}=1$ theories from being lifted by quantum corrections by time-reversal symmetries is discussed in \cite{Gaiotto:2018yjh}.

\subsection{Review of double-scaling limits at large charge}

We now review the double-scaling limits used to compute observables in the large charge expansion. Our review will be very brief, and for more details we refer the reader to the original papers cited below.

\subsubsection{$\epsilon$-expansion}

In perturbative examples, the large-charge limit is organized by a double-scaling limit \cite{Arias-Tamargo:2019xld,Badel:2019khk,Watanabe:2019pdh} (we will also need to include fermions, see \cite{Sharon:2020mjs}). The reason is that the classical fixed-charge saddle depends on the charge $Q$ and the coupling $g$ only through the combination
\begin{equation}
  x\equiv g^2Q.
\end{equation}
One therefore takes
\begin{equation}
  Q\gg 1, \qquad g\ll 1, \qquad x=g^2Q\ \text{is fixed}.
\end{equation}
In this regime the lowest operator dimension admits an expansion of the form
\begin{equation}
  \Dm(Q,g) = \frac{1}{g^2}\Delta_{-1}(x) + \Delta_0(x) + g^2\Delta_1(x)+\cdots, \qquad x=g^2Q .
\end{equation}
The leading term is obtained by solving the classical fixed-charge equations, while $\Delta_0(x)$ is the one-loop correction from the quadratic fluctuations around the helical saddle on $\mathbb R\times S^{d-1}$.

Thus the computation naturally splits into a classical and a quantum piece. First one finds the homogeneous fixed-charge saddle of the classical action at generic coupling and evaluates the cylinder energy on that saddle. This gives
\begin{equation}
  \Delta_{\rm cl}(Q,g) = \frac{1}{g^2}\Delta_{-1}(x).
\end{equation}
One then expands the fields around the helical background,
\begin{equation}
  \phi=\phi_*+\delta\phi
\end{equation}
and computes the Gaussian determinant of quadratic fluctuations. This gives the first quantum correction,
\begin{equation}
  \Delta_{\rm 1-loop}(Q,g) = \Delta_0(x).
\end{equation}
so that, at fixed $x=g^2Q$,
\begin{equation}
  \Dm(Q,g) = \Delta_{\rm cl}(Q,g)+\Delta_{\rm 1-loop}(Q,g)+O(1/Q).
\end{equation}
This is the only general input from the perturbative large-charge formalism that we will need later on.

Let us mention one subtlety in applying this logic to $3d$ $\mathcal N=1$ theories. Since minimal supersymmetry in $4d$ has four real supercharges, it is not clear how to write a theory in $4-\epsilon$ dimensions which interpolates between a 4d theory and a 3d theory with 2 supercharges. For Wess-Zumino models, this difficulty can be bypassed by embedding the model in a non-supersymmetric Yukawa theory and analytically continuing the number of fermionic degrees of freedom to the value corresponding to a single $3d$ Majorana fermion \cite{Fei:2016sgs,ThomasSeminar}.  It is less clear whether an analogous continuation works for the gauge theories discussed in this paper, and so we will not use it there; see \cite{Breitstein:2024pwi} for a study of this question in $3d$ $\mathcal N=1$ SQED.

\subsubsection{Large $N$}

There is a second useful regime in which the microscopic theory is strongly coupled but admits a controlled large-$N$ description. The prototype is the large-$N$ treatment of the $O(N)$ model at fixed charge \cite{Alvarez-Gaume:2019biu,Giombi:2020enj} (see also the review \cite{Gaume:2020bmp}). The calculations below follow the same logic, adapted to theories with moduli spaces.

The starting point is the canonical partition function at fixed charge. For a single $U(1)$ charge one projects the grand-canonical partition function onto charge $Q$ by introducing a chemical potential $\theta$ conjugate to the charge:
\begin{equation}
    Z_Q(\beta) =
    \int_0^{2\pi}\frac{d\theta}{2\pi}
    e^{-i\theta Q}
    Z(i\theta)
    = \int_0^{2\pi}\frac{d\theta}{2\pi}
    e^{-i\theta Q}
    \Tr\,e^{-\beta H+i\theta \hat Q}.
\end{equation}
For several commuting Cartans one similarly introduces chemical potentials $\theta_a$ conjugate to charges $Q_a$,
\begin{equation}\label{eq:Z_chem}
  Z_{\{Q_a\}}(\beta) = \int_0^{2\pi}\prod_a\frac{d\theta_a}{2\pi} e^{-i\sum_a\theta_aQ_a} \Tr\,e^{-\beta H+i\sum_a\theta_a\hat Q_a}.
\end{equation}
At low temperature we expect
\begin{equation}
  Z_{\{Q_a\}}(\beta) \approx e^{-\beta\Dm(Q_a)} \qquad (\beta\to\infty),
\end{equation}
so the saddle-point evaluation of \eqref{eq:Z_chem} directly computes the lowest operator dimension.

In a path-integral description, the $\theta_a$ may be viewed as constant background gauge fields along the Euclidean time circle, so they enter through the covariant derivative
\begin{equation}
  D_0 = \partial_0+i\theta_aT^a
\end{equation}
on fields charged under the Cartan generators $T^a$. One then rewrites the interacting theory using standard Hubbard-Stratonovich fields, integrates out the $N$ matter fields, and obtains an effective action for a small number of collective fields. Schematically, at leading order in $N$ we find
\begin{equation}
  Z_{\{Q_a\}}(\beta) = \int d\theta_a\int D\sigma D\phi_0 e^{-N S_{\rm eff}[\theta_a,\sigma,\phi_0]}.
\end{equation}
Here $\phi_0$ denotes possible condensates while $\sigma$ stands for the Hubbard-Stratonovich fields. The whole point of the large-$N$ method is that the coefficient of the effective action is of order $N$, so it can be evaluated by a saddle point approximation. The large charge enters at the same time through the factor $e^{-i\sum_a\theta_aQ_a}$ in the fixed-charge projection. Thus, in the regime relevant for this paper, $1\ll N\ll Q_a$, both the collective-field effective action and the fixed-charge projection are sharply peaked, and the dominant contribution comes from a homogeneous helical saddle.

The saddle-point equations are obtained by varying the action with respect to the collective fields and the chemical potentials,
\begin{equation}
    \frac{\partial S_{\rm eff}}{\partial \theta_a}=0,
    \qquad
    \frac{\delta S_{\rm eff}}{\delta \sigma}=0,
    \qquad
    \frac{\partial S_{\rm eff}}{\partial \phi_0}=0.
\end{equation}
The corresponding grand-canonical free energy is
$F(\theta_a)\equiv-\beta^{-1}\log Z(i\theta_a)$, and the saddle-point condition for the chemical potentials becomes
\begin{equation}
  \frac{\partial F}{\partial(i\theta_a)}=Q_a.
\end{equation}
The lowest operator dimension is then obtained from the Legendre transform
\begin{equation}
  \Dm(Q_a) = \lim_{\beta\to\infty}\left[F(\theta_a)-i\sum_a\theta_aQ_a\right]_{\rm saddle}.
\end{equation}
This also makes clear how the classical and quantum pieces arise. First one finds the homogeneous saddle and evaluates the effective action on it; this gives the leading contribution, which in the examples below determines $\alpha_1$. One then expands the collective fields around that saddle,
\begin{equation}
  \sigma=\sigma_*+\delta\sigma, \qquad \phi_0=\phi_{0,*}+\delta\phi_0
\end{equation}
and computes the Gaussian determinant. This gives the first quantum correction, which in our case contributes to $\alpha_0$. 

\subsection{$SO(N)\times SO(N)$ model}\label{sec:SO(N)SO(N)_epsilon}

We now study a 3d $\mathcal{N}=1$ theory with real superfields $A,X_i,Y_i$, with $i=1,\ldots,N$, and superpotential
\begin{equation}\label{eq:SON_model_W}
  W=\frac{g}{2}A\left(X_{i}^{2}-Y_{i}^{2}\right).
\end{equation}
The model has continuous global symmetry $SO(N)\times SO(N)$. It also has a discrete symmetry $X_i\leftrightarrow Y_i$, which protects the moduli space branch where $\langle |X_i|^2\rangle=\langle |Y_i|^2\rangle$, and a second discrete symmetry $A\to -A$, which protects the moduli space branch with $\langle A\rangle\neq 0$. We will focus on the former, where a continuous symmetry is broken. For $N=2$ this model reduces to the 5-field example of \cite{Cuomo:2024fuy}.

In this section we compute the large-charge spectrum of this model in two controlled limits: the $\epsilon$-expansion near 4d, and the large-$N$ limit in the regime $Q_{x,y}\gg N\gg 1$. In appendix~\ref{app:equal-scaling} we generalize the large-$N$ analysis to the complementary equal-scaling regime $Q_{x,y}=\lambda_{x,y}N$, in which the charges grow linearly with $N$.

\subsubsection{$\epsilon$-expansion}

We apply the general double-scaling discussion to the model \eqref{eq:SON_model_W}. We will focus on turning on a single Cartan of each $SO(N)$ symmetry, so we choose two components of $X_i$ and combine them into a complex field $x$, and similarly choose two components of $Y_i$ and combine them into a complex field $y$. It turns out that in this case the answer follows almost immediately from the result for $N=2$ computed in \cite{Cuomo:2024fuy}. Indeed, writing the remaining real fields as $X_I,Y_I$, with $I=1,\ldots,N-2$, the superpotential becomes
\begin{equation}
  W=\frac{g}{2}A\left(|x|^2-|y|^2+X_I^2-Y_I^2\right).
\end{equation}
The charged sector then has a $U(1)_x\times U(1)_y$ symmetry rotating $x$ and $y$. For generic $Q_x,Q_y$, only the $(A,x,y)$ sector participates in the charged saddle, and so the result reduces to that of the $N=2$ case. In the large-charge regime $g_*^2\sqrt{Q_x^2+Q_y^2}\gg1$, we find
\begin{equation}
    \Delta_{\rm cl}(Q_x,Q_y) = \sqrt{2}\left(1-\frac{\epsilon}{2} +\frac{\epsilon}{4(N+4)}
    +O(\epsilon^2)\right) \sqrt{Q_x^2+Q_y^2}.
    \label{eq:son-epsilon-alpha1-general}
\end{equation}
From this result one can read off $\alpha_1$ in any direction in charge space.

To find $\alpha_0$, it is convenient to restrict to the diagonal charge assignment
\begin{equation}
  Q_x=Q_y\equiv Q.
\end{equation}
To obtain the next correction, we expand around this saddle and compute the Casimir energy of the quadratic fluctuations. The key point is that the nontrivial interacting sector is again only the $(A,x,y)$ sector, exactly as in the $N=2$ 5-field model computed in \cite{Cuomo:2024fuy}. The additional $X_I,Y_I$ fields are free on this background, so at this order they only affect the answer through the fixed-point value of the coupling. It follows that the one-loop corrected dimension takes the form
\begin{equation}
  \Dm(Q)=\left(d-2+\frac{g^2}{64\pi^2}+O(g^4)\right)Q-\left(\frac{1}{32}+O(g^2)\right)\log(g^2Q)+O(g^0Q^0).
\end{equation}
The one-loop beta function can be computed explicitly and is
\begin{equation}
  \beta_g=-\frac{g\epsilon}{2}+\frac{g^3(N+4)}{64\pi^2},
\end{equation}
Therefore the fixed point is at
\begin{equation}
  g_*^2=\frac{32\pi^2}{N+4}\epsilon.
\end{equation}
Substituting $g=g_*$ gives
\begin{equation}
  \Dm(Q,Q) =\left(2+\left(\frac{1}{2(N+4)}-1\right)\epsilon +O(\epsilon^2)\right)Q -\frac{1}{32}\log(\epsilon Q)+O(Q^0).
\end{equation}
And so for the diagonal charge assignment $Q_x=Q_y$,
\begin{equation}
  \alpha_1 = 2+ \left(\frac{1}{2(N+4)}-1\right)\epsilon +O(\epsilon^2), \qquad \alpha_0 = -\frac{1}{32}+O(\epsilon).
\end{equation}

\subsubsection{Large $N$}

We now analyze the same model directly in the large-$N$ expansion in 3d. The logic is parallel to the general discussion above: we first determine the homogeneous fixed-charge saddle and then evaluate the one-loop determinants around it. Expanding the superfields as
\begin{gather}
A(x,\theta) = a + \bar{\theta} \eta + \frac12 \bar{\theta}\theta F, \notag\\
X_i(x,\theta) = X_i + \bar{\theta} \, \psi_{X_i} + \frac12 \bar{\theta}\theta F_{X_i}
\end{gather}
and similarly for $Y_i$, we find the component action
\begin{gather}
    S = \frac12(\partial_\mu X_i)^2 + \frac12 (\partial_\mu Y_i)^2 + i \psi_{X_i}\gamma^\mu \partial_\mu \psi_{X_i} + i \psi_{Y_i}\gamma^\mu \partial_\mu \psi_{Y_i} + F_{X_i}^2 + F_{Y_i}^2 + \notag\\
    +2 a (X_i F^X_i - Y_i F^Y_i) - a (\bar{\psi}^X_i \psi^X_i - \bar{\psi}^Y_i \psi^Y_i) + F(X_i^2 - Y_i^2) - 2 \bar{\eta}(X_i \psi^X_i - Y_i \psi^Y_i).
\end{gather}
Here $F$ already plays the role of the Hubbard-Stratonovich field, so no additional auxiliary field is needed. This implies that at the IR fixed point $\Delta_F=2$, and supersymmetry then fixes $\Delta_A=1$.

To construct the large-$N$ effective action we allow homogeneous expectation values for the charged combinations $X\equiv X_1+iX_2$ and $Y\equiv Y_1+iY_2$. We work on $\mathbb R\times S^2$, so the scalars acquire the usual conformal mass term $\frac14(X_i^2+Y_i^2)$.\footnote{Since $\Delta_A=1$, no conformal mass term is added for $a$.} Introducing chemical potentials $\theta_x,\theta_y$ for the two Cartans with charges $Q_x,Q_y$, we obtain
\begin{gather}
    S_{\rm eff} = \frac{N}{2} \operatorname{tr}\log \left[-\Delta + F - a^2 + \frac14\right] + \frac{N}{2} \operatorname{tr}\log \left[-\Delta - F - a^2 + \frac14 \right] - \notag\\
    - N \operatorname{tr}\log\left[i \gamma^\mu \partial_\mu + a \right]  - N \operatorname{tr}\log\left[i \gamma^\mu \partial_\mu - a \right] + \notag\\
+    i \theta_x Q_x +  i \theta_y Q_y +  \left(F + \theta_x^2 + \frac14\right) X^2   +  \left(-F + \theta_y^2 + \frac14\right) Y^2 + \notag\\
    + 2 a (XF_x - Y F_y) + F_{x}^2 + F_{y}^2. \label{eq:son-Seff}
\end{gather}
Here $\Delta$ is the Laplacian on $\mathbb R\times S^2$. In the regime $Q_{x,y}\gg N\gg 1$, the classical charged condensates contribute at order $Q$ while the $\Tr\log$ terms contribute at order $N$, so the determinants do not modify the leading saddle-point equations, though they do contribute to the first subleading correction. At leading order we therefore find the saddle-point equations
\begin{gather}
\left\{
    \begin{matrix}
       2 a X + 2 F_{x} &= 0, \\
       -2 a Y + 2 F_y &= 0,\\
        2\left(F + \theta_x^2 + \frac14\right) X  +2 a F_x &= 0,\\
       2\left(-F + \theta_y^2 + \frac14\right) Y   - 2 a F_y &= 0,\\
       X F_x - Y F_y &= 0,\\
     X^2 - Y^2 &= 0, \\
     i Q_x + 2 \theta_x X^2 = i Q_y + 2\theta_y Y^2 &= 0
    \end{matrix}
\right.
\end{gather}
From the first two and the fifth equations we immediately find $F_x=F_y=a=0$, while the sixth equation gives $Y=X$. The remaining saddle-point equations reduce to
\begin{gather}
\left\{
    \begin{matrix}
        2\left(F + \theta_x^2 + \frac14\right) X = 0,\\
       2\left(-F + \theta_y^2 + \frac14\right) X  = 0,\\
     i Q_x + 2 \theta_x X^2 = i Q_y + 2\theta_y X^2 = 0
    \end{matrix}
\right.
\end{gather}
whose solution is
\begin{gather}
    \theta_{x,y} = -\frac{i Q_{x,y}}{\sqrt{2(Q_x^2 + Q_y^2)}}, \notag\\
    X^2 =Y^2= \sqrt{Q_x^2+ Q_y^2},\quad  
    F = \frac{Q_x^2 - Q_y^2}{4(Q_x^2+ Q_y^2)}. \label{eq:son-saddle}
\end{gather}
Evaluating the classical action on this saddle gives the leading contribution
\begin{gather}
    \Dm(Q_x,Q_y) = \sqrt{\frac{Q_x^2 + Q_y^2}{2}}. \label{eq:son-Delta-leading}
\end{gather}

\begin{figure}     \centering
\includegraphics[width=0.5\linewidth]{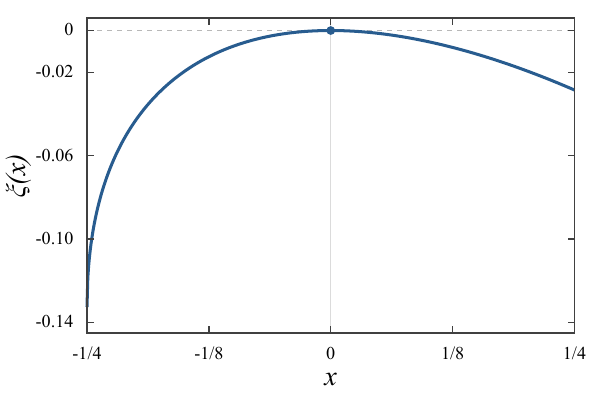}
     \caption{$\xi(x)$ for $-\frac14\leq x\leq \frac14$. The function vanishes at $x=0$ and is negative everywhere else.}
     \label{fig:xi}
 \end{figure}

We now turn to the subleading contribution. This comes from evaluating the $\Tr\log$ terms on the leading saddle. It is useful to define the regulated determinant
\begin{gather}\label{eq:xi}
    \xi(x) \equiv \frac{1}{T}\operatorname{tr} \log\left[-\Delta + x + \frac14\right] = \sum_l \left[\left(l+\frac12\right)\sqrt{\left(l + \frac12\right)^2 + x} - \left(l+\frac12\right)^2 - \frac{x}{2} \right].
\end{gather}
We have regulated the determinant by subtracting the first two terms in the large-$\ell$ expansion,
\begin{equation}
  r\sqrt{r^2+x}=r^2+\frac{x}{2}+O(r^{-2}), \qquad r=\ell+\frac12.
\end{equation}
Equivalently, this is dimensional regularization with local power divergences removed. With this convention a conformally coupled real scalar on $\mathbb R\times S^2$ has vanishing Casimir energy, $\xi(0)=0$. We plot $\xi(x)$ for the relevant range $-\frac14\leq x\leq \frac14$ in figure \ref{fig:xi}.

The final result is
\begin{gather}
\Dm(Q_x,Q_y) =  \sqrt{\frac{Q_x^2 + Q_y^2}{2}} + \frac{N}{2}\xi\left(\frac{Q_x^2-Q_y^2}{4(Q_x^2 +Q_y^2)}\right) + \frac{N}{2}\xi\left(\frac{Q_y^2-Q_x^2}{4(Q_x^2 +Q_y^2)}\right) + \cdots. \label{eq:son-Delta-final}
\end{gather}
There are additional corrections of order $N^0Q^0$ which we have not computed, as well as subleading orders in $1/Q$. We have thus found that the leading linear term in $Q$ (which determines $\alpha_1$) is
\begin{equation}
  \sqrt{\frac{Q_x^2 + Q_y^2}{2}},
\end{equation}
while the order-$NQ^0$ correction is
\begin{equation}
    \alpha_0=\frac{N}{2}\xi\left(\frac{Q_x^2-Q_y^2}{4(Q_x^2 +Q_y^2)}\right) + \frac{N}{2}\xi\left(\frac{Q_y^2-Q_x^2}{4(Q_x^2 +Q_y^2)}\right).
\end{equation}
It is clear from figure \ref{fig:xi} that $\alpha_0$ is always non-positive.

\subsection{SO(3) model}

We now consider a 3d $\mathcal N=1$ Wess-Zumino model with three real superfields $x_i,y_i,z_i$, $i=1,2,3$, and superpotential
\begin{equation}
  W = g\,\epsilon_{ijk}x_i y_j z_k .
\end{equation}
We will work in the $\epsilon$-expansion in $d=4-\epsilon$ dimensions. First, a direct computation shows that the theory has beta function
\begin{equation}
  \beta_g = -\frac{\epsilon}{2}g + \frac{g^3}{8\pi^2} +O(g^5),
\end{equation}
and so a fixed point exists in the $\epsilon$-expansion.

The model has an $SO(3)$ global symmetry. The F-term equations are
\begin{equation}
  x \wedge y = y \wedge z = z \wedge x =0 ,
\end{equation}
so on the classical moduli space the three vectors are parallel, and the discrete symmetries $x\to -x$, $y\to -y$ and $z\to -z$ protect this moduli space from being lifted by quantum corrections. We may therefore write
\begin{equation}
  x_i = r\,s_1 n_i,\qquad y_i = r\,s_2 n_i,\qquad z_i = r\,s_3 n_i ,
\end{equation}
where $n_i n_i=1$ and $s_a s_a=1$. The moduli space is locally
\begin{equation}
  \mathcal M \simeq \mathbb R_+ \times S^2_n \times S^2_s .
\end{equation}

We fix a charge $Q$ under the Cartan of the $SO(3)$ symmetry. We can choose the homogeneous fixed-charge saddle to rotate along a great circle of $S^2_n$, while $s$ is constant:
\begin{equation}
  r=v,\qquad n(t)=(\cos \mu t,\sin \mu t,0),\qquad s=s_0 .
\end{equation}
On the unit-radius cylinder the conformal mass is $ m_d=\frac{d-2}{2}$, and the equation of motion for $r$ fixes $ \mu=m_d$. The charge-fixing condition is
\begin{equation}
  Q=\Omega_{d-1} v^2 \mu .
\end{equation}
Evaluating the classical energy on the saddle gives
\begin{equation}
  \Delta_{\rm cl}(Q)=\mu Q = \frac{d-2}{2}Q,
\end{equation}
so that $\alpha_1$ is just the free result: 
\begin{equation}
  \alpha_1 = \frac{d-2}{2}.
\end{equation}

We now discuss the one-loop correction. The charged sector $(r,S^2_n)$ is precisely the $CP^1\simeq S^2$ sigma-model sector with the normalization $c=1$ in the notation of \cite{Cuomo:2024fuy}. Therefore its fluctuation determinant is the same as the $CP^1$ determinant computed there. Equivalently, this sector is just the free conformal scalar sector written in polar variables, and its regulated Casimir energy vanishes:
\begin{equation}
  \Delta^{(r,n)}_{\rm 1-loop}=0 .
\end{equation}

The remaining factor $S^2_s$ gives two neutral angular moduli. Expanding around a point on $S^2_s$,
\begin{equation}
  s=(\eta_1,\eta_2,\sqrt{1-\eta_1^2-\eta_2^2}) ,
\end{equation}
the quadratic action is
\begin{equation}
  \mathcal L^{(2)}_s = \frac{v^2}{2} \sum_{a=1}^2 \partial_\mu \eta_a \partial^\mu \eta_a .
\end{equation}
Thus the two $\eta_a$ have the same dispersion relation as the neutral spectator modulus in the five-field model of \cite{Cuomo:2024fuy},
\begin{equation}
  \omega_s(\ell)=\sqrt{\ell(\ell+d-2)} .
\end{equation}
We may therefore import the corresponding Casimir result directly. In $d=4-\epsilon$, one such real spectator mode contributes
\begin{equation}
  -\frac{1}{32}\log(\epsilon Q)
\end{equation}
to $\Dm(Q)$. Since the $SO(3)$ model has two such modes, while the $(r,S^2_n)$ sector contributes zero, we obtain
\begin{equation}
  \Delta_{\rm 1-loop}(Q) = -\frac{1}{16}\log(\epsilon Q)+O(Q^0).
\end{equation}
The fermions do not modify this logarithmic term: after diagonalizing the fermion mass matrix in the rotating frame, their contribution reduces to the shifted fermion towers analyzed in \cite{Cuomo:2024fuy}, whose regulated logarithmic contribution vanishes.

Combining the classical and one-loop pieces gives
\begin{equation}
  \Dm(Q) = \left(1-\frac{\epsilon}{2}\right)Q -\frac{1}{16}\log(\epsilon Q) +O(Q^0) .
\end{equation}
Thus, in the notation of \eqref{eq:intro-large-charge-4d}, this example has
\begin{equation}
  \alpha_1=1-\frac{\epsilon}{2}, \qquad \alpha_0=-\frac{1}{16}
\end{equation}
at this order in $\epsilon$.

\subsection{SQED}\label{sec:SQED}

Next we discuss 3d $\mathcal{N}=1$ SQED at large $N_f$. Suppressing the gauge kinetic term, the  Lagrangian contains
\begin{equation}
|D\phi_i|^2+i\bar\psi_i\cancel{D}\psi_i+\frac{i}{2}\bar\lambda\cancel{D}\lambda-i\bar\psi_i\lambda\phi_i+i\bar\lambda\psi_i\bar\phi_i,\quad i=1,\ldots,N_f.
\end{equation}
Since we are working on $\mathbb{R}\times S^2$, we must also include the conformal mass term
\begin{equation}
  m^2|\phi_i|^2,\quad m^2=1/4\;.
\end{equation}
The theory has a global $SU(N_f)$ symmetry. Its moduli space is obtained by giving arbitrary expectation values to $\phi_i$, modulo the overall $U(1)$ gauge symmetry. Thus the moduli-space EFT consists of a dilaton, parametrizing the radial direction, coupled to a nonlinear sigma model with target space $\mathbb{CP}^{N_f-1}$. This moduli space is preserved to all orders by the time-reversal symmetry $\phi_i\to -\phi_i$ \cite{Gaiotto:2018yjh}.

To project onto fixed charges under the Cartan of $SU(N_f)$ we introduce chemical potentials $\theta_i$, together with a Lagrange multiplier $\lambda$ imposing $\sum\limits_i\theta_i=0$. This amounts to adding
\begin{equation}
  -i\theta_i Q_i + i\lambda \sum_i\theta_i.
\end{equation}
At the order in $N_f$ relevant for our discussion we may ignore the gaugino and gauge-field fluctuations. To obtain the large-$N_f$ effective action on $\mathbb R\times S^2$, we keep the homogeneous charged scalar condensates $\phi_i$, the constant gauge-field mode $A_0$, and the chemical potentials $\theta_i$, and integrate out the nonzero modes of the matter multiplets. The resulting effective action is
\begin{equation}
\begin{split}
    \frac{S_{\rm eff}}{T}
    =& \sum_i\Bigl[
    \bigl((A_0+\theta_i)^2+m^2\bigr)|\phi_i|^2
    -i\theta_iQ_i
    \Bigr] +i\lambda\sum_i\theta_i
    \\
    &+\sum_i\Tr\log\!\left[-D_{0,i}^2-\nabla^2+m^2\right]
    - \sum_i\Tr\log\!\left[i\cancel{D}_{0,i}\right]+O(N_f^0),
\end{split}
\end{equation}
where
\begin{equation}
  D_{0,i}=\partial_0+i(A_0+\theta_i).
\end{equation}
In the regime $Q_i\gg N_f\gg 1$ the $\Tr\log$ terms do not modify the leading saddle-point equations, so at leading order we may extremize only the classical part of $S_{\rm eff}$. This gives
\begin{gather}
\left\{
    \begin{matrix}
        \left((A_0+\theta_i)^2+m^2\right)\phi_i &= 0,\\
        2(A_0+\theta_i)|\phi_i|^2-iQ_i+i\lambda &= 0,\\
        \sum_i (A_0+\theta_i)|\phi_i|^2 &= 0,\\
        \sum_i\theta_i &= 0.
    \end{matrix}
\right.
\end{gather}
Summing the second equation over $i$ and using the third one gives
\begin{equation}
  \lambda = \frac{1}{N_f}\sum_i Q_i.
\end{equation}
For generic charges all $\phi_i$ are nonzero, since otherwise the second equation would impose a nongeneric relation among the $Q_i$. The first equation then implies
\begin{equation}
  A_0+\theta_i=i s_i m,\qquad s_i\in\{\pm1\}.
\end{equation}
Imposing $\sum_i\theta_i=0$ fixes
\begin{equation}
  A_0=\frac{im}{N_f}\sum_i s_i.
\end{equation}
The second equation then determines
\begin{equation}
  s_i=\operatorname{sgn}\left(Q_i-\frac{1}{N_f}\sum_jQ_j\right).
\end{equation}
Substituting back into the classical action gives
\begin{equation}
  \Delta=-i\sum_i\theta_iQ_i =m\sum_i \operatorname{sgn}\left(Q_i-\frac{1}{N_f}\sum_jQ_j\right)Q_i.
\end{equation}
It is convenient to define the shifted charges
\begin{equation}
  \tilde Q_i = Q_i-\frac{1}{N_f}\sum_jQ_j.
\end{equation}
Then the leading contribution becomes
\begin{equation}
  \Delta = \frac12\sum_i|\tilde Q_i|,
\end{equation}
which is just the free-theory result.

We now turn to the subleading $N_fQ^0$ correction. This is obtained by evaluating the $\Tr\log$ terms on the leading saddle. On the cylinder the constant background values of $A_0$ and $\theta_i$ can be absorbed into the time dependence of the charged fields, so the bosonic determinant reduces to the free conformally coupled scalar result $\sum_i\Tr\log(-\partial^2+m^2)$ while the fermionic determinant reduces to the free fermion result. Their Casimir-energy contribution therefore vanishes. We thus find
\begin{equation}
  \Dm(Q)=\frac12 \sum_i|\tilde{Q}_i|+O(N_f^0Q^0)+O(1/Q).
\end{equation}
In particular, we find that $\alpha_0$ vanishes at leading order in $1/N_f$. The result is consistent with the EFT computation in \cite{Cuomo:2024fuy}; taking $N_f\to \infty$ in \cite{Cuomo:2024fuy} and using the fact that the Wilson coefficient $c$ there approaches the free value $1/4$ as $N_f\to\infty$, we find that the result indeed reduces to the Casimir energy of a scalar with only the conformal coupling on $\mathbb{R}\times S^2$, which vanishes.

\subsection{$SU(2)$ SQCD}\label{sec:SU2_SQCD}

We now consider the Higgs branch of 3d $\mathcal N=1$ $SU(2)$ SQCD with $N_f$ complex fundamental multiplets, following~\cite{Gaiotto:2018yjh}. Since the $SU(2)$ fundamental is pseudoreal, the scalar quarks can be combined into an $N_f$-component quaternionic vector,
\begin{equation}
  q\in \mathbb H^{N_f}.
\end{equation}
The gauge group is $SU(2)_{\rm gauge}\simeq Sp(1)$, acting by left quaternionic multiplication, and the flavor symmetry is enhanced from the generic $U(N_f)$ to $  Sp(N_f)$. On the moduli space, the classical quotient is therefore
\begin{equation}
  \mathcal M_{\rm reg} = \left(\mathbb H^{N_f}\setminus\{0\}\right)/Sp(1) = \mathbb R_+\times \mathbb HP^{N_f-1}.
\end{equation}
Thus the low-energy theory consists of the dilaton together with a NLSM with target space
\begin{equation}
  \mathbb HP^{N_f-1} = \frac{Sp(N_f)}{Sp(N_f-1)\times Sp(1)}.
\end{equation}

For a general $SU(N_c)$ gauge group the structure of the Higgs branch can be more involved - for higher $N_c$ there are separate meson and baryon branches, and the NLSM contains moduli which are not all Goldstone bosons, making an equally explicit analysis less immediate. In addition, for $N_f<N_c$ the moduli space can be lifted nonperturbatively. We will thus focus on the $N_c=2$ case, where the moduli space has this particularly simple quotient description.

\subsubsection{Two-derivative EFT and $\alpha_1$}

The bosonic two-derivative EFT for the moduli space is
\begin{equation}
  \mathcal L_{\rm EFT}
  =
  \frac12(\partial\Phi)^2
  -\frac12 m_d^2\Phi^2
  +\frac c2\Phi^2 g_{AB}(x)\partial_\mu x^A\partial^\mu x^B
  +\cdots,
  \label{eq:sqcd-eft}
\end{equation}
where $\Phi$ is the dilaton, $x^A\in \mathbb HP^{N_f-1}$ and $g_{AB}$ is the quotient metric on $\mathbb HP^{N_f-1}$. $c>0$ is a Wilson coefficient whose exact value we do not know, but we will be able to compute it at large $N_f$ later on. In 3d on the unit cylinder $\mathbb R\times S^2$, $m_d=1/2$. All computations will be performed using dimensional regularization, which allows us to preserve the symmetries of the theory at the quantum level.

We choose a charge assignment with a single nonzero Cartan charge $Q$ for a $U(1)\subset Sp(N_f)$. Let $k^A(x)$ be the Killing vector on $\mathbb HP^{N_f-1}$ generated by this $U(1)$. For the rank-one charge
assignment considered here, the orbit lies inside a totally geodesic
\begin{equation}
  \mathbb HP^1\subset \mathbb HP^{N_f-1}.
\end{equation}
With the quotient metric inherited from $S^{4N_f-1}\to\mathbb HP^{N_f-1}$, this $\mathbb HP^1$ is a round
four-sphere of radius $a=\frac12$. The fixed-charge saddle sits at a point where $|k|^2=g_{AB}k^Ak^B$ is maximal, and for this $U(1)$ action
\begin{equation}
  k_{\max}^2=a^2=\frac14.
\end{equation}

Writing the helical saddle as
\begin{equation}
  \Phi=v, \qquad \dot x^A=\mu k^A(x_0),
\end{equation}
the dilaton equation of motion gives
\begin{equation}
  m_d^2=c\mu^2 k_{\max}^2, \qquad\Rightarrow\qquad \mu=\frac1{\sqrt c}.
\end{equation}
The classical fixed-charge energy is therefore
\begin{equation}
  \Delta_{\rm cl}(Q)=\mu Q=\frac{Q}{\sqrt c},
\end{equation}
so that
\begin{equation}
  \alpha_1=\frac1{\sqrt c}.
\end{equation}

\subsubsection{Scalar Casimir contribution}

We now compute the scalar contribution to $\alpha_0$ from the quadratic fluctuations around this saddle. Let
\begin{equation}
  \lambda_\ell=\ell(\ell+1), \qquad \ell=0,1,2,\ldots,
\end{equation}
be the scalar spherical-harmonic eigenvalues on $S^2$. We expand the fields as
\begin{equation}
    \Phi=v+\delta\Phi,
    \qquad
    x^A=x_s^A(t)+\delta x^A,
    \qquad
    \dot x_s^A=\mu k^A(x_s),
\end{equation}
and decompose the target-space fluctuation into the component tangent to the helical orbit and the components orthogonal to it,
\begin{equation}
    \delta x^A
    =
    \delta x_\parallel\,\hat k^A(x_s)
    +
    \delta x_\perp^A,
    \qquad
    \hat k^A(x_s)=\frac{k^A(x_s)}{k_{\max}},
    \qquad
    k_A(x_s)\delta x_\perp^A=0.
\end{equation}
First consider the two fluctuations consisting of the radial mode $\delta\Phi$ and the angular fluctuation tangent to the helical orbit $\delta x_\parallel$. Their two frequencies are
\begin{equation}
  \omega_{\ell,-}=\ell, \qquad \omega_{\ell,+}=\ell+1.
\end{equation}
The $\ell=0$ mode in the lower branch is the collective coordinate associated with the broken $U(1)$. The regulated zero-point energy of this block is
\begin{equation}
    \frac12\sum_{\ell=0}^{\infty}
    (2\ell+1)\left(\omega_{\ell,+}+\omega_{\ell,-}\right)
    =
    \frac12\sum_{\ell=0}^{\infty}(2\ell+1)^2
    =0
\end{equation}
in dimensional regularization. We therefore only need the remaining fluctuations orthogonal to the helical orbit $\delta x_\perp$.

Inside the selected $\mathbb HP^1$, there are three real transverse modes, which are
\begin{align}
  \omega_{\ell,0}
  &=
  \sqrt{\lambda_\ell+\frac1c},
  \\
  \omega_{\ell,\pm}
  &=
  \sqrt{\lambda_\ell+\frac1c}
  \pm\frac1{\sqrt c}.
\end{align}
so the entire $\mathbb HP^1$ block contributes 
\begin{equation}
  \omega_{\ell,0}+\omega_{\ell,+}+\omega_{\ell,-} = 3\sqrt{\lambda_\ell+\frac1c}.
\end{equation}
The remaining $4(N_f-2)$ real modes are normal to the selected $\mathbb HP^1$. They organize into $2(N_f-2)$ pairs
\begin{equation}
  \omega_{\ell,+}^{\perp}+\omega_{\ell,-}^{\perp} = 2\sqrt{\lambda_\ell+\frac1{4c}}.
\end{equation}

To express the answer in the notation already used above, recall the regulated scalar determinant function $\xi$ defined in \eqref{eq:xi}. With this convention, $\xi(y)$ is the Casimir sum for a scalar with frequency $\sqrt{\lambda_\ell+y+1/4}$. Since
\begin{equation}
  \ell(\ell+1)+x=\left(\ell+\frac12\right)^2+x-\frac14,
\end{equation}
a real relativistic scalar with frequency
\begin{equation}
  \omega_\ell=\sqrt{\lambda_\ell+x}
\end{equation}
contributes
\begin{equation}
  \frac12 \sum_{\ell=0}^{\infty} (2\ell+1)\sqrt{\ell(\ell+1)+x}\Big|_{\rm reg} = \xi\left(x-\frac14\right). \label{eq:sqcd-xi-shift}
\end{equation}
Therefore the scalar contribution to the order-$Q^0$ term is
\begin{equation}
  \alpha_0^{\rm scalar} = 3\xi\left(\frac1c-\frac14\right) + 4(N_f-2)\xi\left(\frac1{4c}-\frac14\right). \label{eq:sqcd-alpha0}
\end{equation}
The first term comes from the $\mathbb HP^1\simeq S^4$ sector, and the second from the normal modes. Thus the scalar contribution is non-positive for $c>0$:
\begin{equation}
  \alpha_0^{\rm scalar}\leq0.
\end{equation}

\subsubsection{Fermion Casimir contribution}

We now include the fermions in the two-derivative moduli-space EFT. In a 3d $\mathcal N=1$ sigma model, their
two-derivative action is fixed by the same metric that appears in the bosonic EFT. In the present case the target-space metric is
\begin{equation}
  ds^2_{\mathcal M} = d\Phi^2 + c\Phi^2 g_{AB}(x)dx^A dx^B, \qquad x^A\in \mathbb HP^{N_f-1}.
\end{equation}
In an orthonormal frame on the target space, the quadratic fermion action around a bosonic background takes the form
\begin{equation}
  \mathcal L_{\rm ferm}^{(2)}
  =
  \frac{i}{2}
  \bar\chi_{\hat I}\gamma^\mu
  \left(
    \nabla_\mu\delta_{\hat I\hat J}
    +
    \partial_\mu Y^K\omega_{K,\hat I\hat J}
  \right)
  \chi_{\hat J}.
\end{equation}
Here $Y^I=(\Phi,x^A)$, hatted indices denote tangent-space indices on $\mathcal M$, and $\omega_{K,\hat I\hat J}$ is the target-space spin connection. On the homogeneous saddle,
\begin{equation}
  \Phi=v, \qquad \dot x^A=\mu k^A(x_0), \qquad \mu=\frac1{\sqrt c},
\end{equation}
the pullback of the target-space spin connection is a constant antisymmetric matrix,
\begin{equation}
  \Omega_{\hat I\hat J} = \mu k^A\omega_{A,\hat I\hat J}\big|_{Y=(v,x_0)} .
\end{equation}
We can diagonalize this antisymmetric matrix into $2\times2$ blocks. Each block can be put into the form
\begin{equation}
  \Omega_{\hat I\hat J}
  =
  \nu
  \begin{pmatrix}
    0 & -1\\
    1 & 0
  \end{pmatrix}
\end{equation}
for some $\nu\geq0$. $\nu$ is the induced fermion chemical potential for that two-Majorana block. Thus each nonzero $2\times2$ block of $\Omega$ gives a pair of Majorana fermions with frequencies shifted by $\pm\nu$.

The tangent space decomposes as
\begin{equation}
  T\mathcal M
  =
  \left(\mathbb R_\Phi\oplus\mathbb R_{\parallel}\right)
  \oplus
  \mathbb R^3_{\mathbb HP^1,\perp}
  \oplus
  \mathbb R^{4(N_f-2)}_{\perp}.
\end{equation}
The first factor consists of the radial fermion and the fermion tangent to the helical orbit. Since
\begin{equation}
  \sqrt c\,\mu\, k_{\max} = \sqrt c\cdot\frac1{\sqrt c}\cdot\frac12 = \frac12,
\end{equation}
this pair gives one two-Majorana block with
\begin{equation}
  \nu=\frac12.
\end{equation}
This is below the first spinor level on $S^2$, and therefore this block does not contribute to the Casimir energy.

Inside the selected $\mathbb HP^1\simeq S^4$, the three transverse fermions split into one unshifted Majorana fermion and one shifted two-Majorana block. The corresponding chemical potentials are
\begin{equation}
  \nu=0
\end{equation}
for the unshifted fermion, and
\begin{equation}
  \nu=\frac1{\sqrt c}
\end{equation}
for the shifted pair. The $4(N_f-2)$ real fermions normal to the selected $\mathbb HP^1$ organize into $2(N_f-2)$ shifted two-Majorana blocks, each with
\begin{equation}
  \nu=\frac1{2\sqrt c}.
\end{equation}

For a two-Majorana block with chemical potential $\nu$, the spinor
frequencies on the unit $S^2$ are
\begin{equation}\label{eq:two-majorana-spectrum}
  \omega_{n,\pm} = n\pm\nu, \qquad n=1,2,\ldots,
\end{equation}
with degeneracy $2n$. In dimensional regularization the contribution of such a fermion block vanishes unless the lower branch crosses zero. Thus there is no contribution for $\nu\leq1$. For $\nu>1$, the negative-energy levels with $n<\nu$ are filled, giving a Fermi sphere contribution
\begin{equation}\label{eq:fermion-casimir-function}
  \mathcal F(\nu) = -2\sum_{n=1}^{\infty}n(\nu-n)_+, \qquad (\nu-n)_+ \equiv \max(\nu-n,0),
\end{equation}
which is manifestly non-positive. Equivalently, if $K=\lfloor\nu\rfloor$, then
\begin{equation}
  \mathcal F(\nu)
  =
  -2\sum_{n=1}^{K}n(\nu-n)
  =
  -\nu K(K+1)
  +
  \frac{K(K+1)(2K+1)}{3}.
\end{equation}
At integer $\nu$, the top level has zero energy, so this expression gives the same Casimir contribution. In particular,
\begin{equation}
  \mathcal F(\nu)=0 \qquad \text{for} \qquad \nu\leq1,
\end{equation}
and
\begin{equation}
  \mathcal F(\nu)\leq0.
\end{equation}
Combining the blocks, the fermion contribution to the order-$Q^0$ term is
\begin{equation}
  \alpha_0^{\rm ferm}(c,N_f) = \mathcal F\left(\frac1{\sqrt c}\right) + 2(N_f-2) \mathcal F\left(\frac1{2\sqrt c}\right).
\end{equation}
The radial/tangent block with $\nu=1/2$ and the unshifted Majorana fermion with $\nu=0$ do not contribute. The total scalar-plus-fermion contribution is therefore
\begin{equation}
  \alpha_0(c,N_f) = \alpha_0^{\rm scalar}(c,N_f) + \alpha_0^{\rm ferm}(c,N_f).
\end{equation}
Using the scalar result above, this gives
\begin{equation}
  \alpha_0(c,N_f)
  =
  3\xi\left(\frac1c-\frac14\right)
  +
  4(N_f-2)\xi\left(\frac1{4c}-\frac14\right)
  +
  \mathcal F\left(\frac1{\sqrt c}\right)
  +
  2(N_f-2)
  \mathcal F\left(\frac1{2\sqrt c}\right).
\end{equation}
Since $\mathcal F(\nu)\leq0$, the fermion contribution is always non-positive. It vanishes whenever the induced chemical potentials remain below the first spinor level on $S^2$. Equivalently, the fermion contribution vanishes for $c\geq 1$; for $\frac14\leq c<1$ only the shifted two-Majorana block inside the selected $\mathbb HP^1$ contributes, while for $c<\frac14$ both this block and the $2(N_f-2)$ normal shifted two-Majorana blocks contribute.

\subsubsection{Large $N_f$ prediction}

At leading order in large $N_f$, the two-derivative metric on the Higgs branch is the classical quotient metric inherited from the canonical kinetic terms of the quarks. Equivalently, writing the quaternionic scalar as a radial
mode times a point in $\mathbb HP^{N_f-1}$, the flat kinetic term descends to $d\Phi^2+\Phi^2 ds^2_{\mathbb HP^{N_f-1}}$. In the normalization of \eqref{eq:sqcd-eft}, this gives
\begin{equation}
  c=1+O(1/N_f).
\end{equation}
At this leading large-$N_f$ value,
\begin{equation}
  \alpha_1=1.
\end{equation}
The normal modes have $x=1/4$, so their scalar Casimir contribution is
\begin{equation}
  \xi(0)=0.
\end{equation}
The remaining $\mathbb HP^1$ block gives
\begin{equation}
  \alpha_0^{\rm scalar} = 3\xi\left(\frac34\right) \simeq -0.556.
\end{equation}
The fermion contribution vanishes at this order. Indeed, at $c=1$ the fermionic chemical potentials are
\begin{equation}
  \nu=1, \qquad \nu=\frac12,
\end{equation}
for the shifted $\mathbb HP^1$ block and the normal shifted blocks, respectively. Since the fermion contribution is controlled by $\mathcal F(\nu)$, which vanishes for $\nu\leq1$, we find
\begin{equation}
  \alpha_0^{\rm ferm}(1,N_f) = \mathcal F(1) + 2(N_f-2)\mathcal F\left(\frac12\right) = 0.
\end{equation}
The $\nu=1$ block has zero modes at the first spinor level, but these do not shift the ground-state energy. Thus the leading large-$N_f$ EFT prediction for the one-Cartan tower is
\begin{equation}
  \Dm(Q) = Q + 3\xi\left(\frac34\right) + O(1/Q).
\end{equation}

\section{Bounds on $\alpha_0$}\label{sec:alpha0}

\subsection{Setup}\label{subsec:alpha0-setup}

As discussed in the introduction, the large-charge expansion for a 3d theory with a single $U(1)$ symmetry broken on the moduli space is
\begin{equation}
  \Dm(Q)=\alpha_1 Q + \alpha_0 + O(1/Q).
\end{equation}
The asymptotic CCC then requires the large-charge convexity bound
\begin{equation}\label{eq:alpha0bound4}
  \alpha_0\leq 0.
\end{equation}
When the global symmetry has several Cartans, however, there are two distinct questions. The first is the fixed-charge-vector problem, where one studies
\begin{equation}
  \Delta_{\min}(Q_a)
\end{equation}
at a fixed charge vector $Q_a$. Fixed-charge-ray convexity would mean applying \eqref{eq:alpha0bound4} to every fixed direction in charge space. The second question is the projected one-charge problem. For a set of Cartans $T^a$ choose a single linear combination
\begin{equation}
  T_n=n_aT^a
\end{equation}
and define
\begin{equation}
  \Delta_{\min}^{(n)}(Q) = \min_{n\cdot Q=Q}\Delta_{\min}(Q_a). \label{eq:projected-one-charge-observable}
\end{equation}
Here only the charge under $U(1)_n$ is held fixed; all orthogonal Cartan charges are minimized over.

The point of this section is that the moduli-space EFT proves the convexity bound for the second type. For the saddle computing $\Delta_{\min}^{(n)}(Q)$, the fermion and gauge-boson contributions are non-positive by the argument from \cite{Cuomo:2024fuy}, and we prove that the scalar contribution to $\alpha_0^{(n)}$ is non-positive as well. Thus the projected observable obeys $\alpha_0^{(n)}\leq0$. When this inequality is strict, the projected tower is asymptotically convex; if it is saturated, the next term in the large-charge expansion controls convexity. Equivalently, for each chosen linear combination $T_n$, the projected minimization selects at least one
charge-space direction on the lower envelope which satisfies the leading convexity bound. This is weaker than proving convexity along every ray in the charge lattice. In fact, we will give a two-derivative $\mathcal{N}=1$ supersymmetric EFT counterexample to the stronger fixed-charge-ray statement. This is not a UV-complete CFT counterexample, but it shows that the stronger claim cannot be derived from the moduli-space EFT alone.

We first recall the part of the moduli-space EFT calculation that we will need. The general setup follows appendix A of \cite{Cuomo:2024fuy}, but we spell out the result in detail. The low-energy theory contains the dilaton $\Phi$, Goldstone coordinates $\pi^a$ for the broken Cartan generators, and additional massless moduli $\phi^A$. Denote the full dimension of the moduli space by $n$. The two-derivative EFT takes the form
\begin{equation}\label{eq:crs-moduli-eft}
\begin{split}
S_{\rm EFT}=&
\int d^d x \sqrt{g}
\Bigg[
\frac{\Phi^2}{2}
\partial_\mu \pi^a \partial^\mu \pi^b
\widehat{G}_{ab}(\phi)
+\Phi^2
\partial_\mu \pi^a \partial^\mu \phi^B
\widehat{G}_{aB}(\phi)
+\\
&+
\frac{\Phi^2}{2}
\partial_\mu \phi^A \partial^\mu \phi^B
\widehat{G}_{AB}(\phi)+
\frac{\widehat{G}_{\Phi\Phi}(\phi)}{2}
\left(
-\Phi \partial^2 \Phi
- m_d^2 \Phi^2
\right)
+\cdots
\Bigg],
\end{split}
\end{equation}
where $m_d=(d-2)/2$ on a unit-radius cylinder. The general homogeneous helical saddle around which we expand is
\begin{equation}\label{eq:helical-saddle-general}
  \Phi=v, \qquad \pi^a=\mu^a t, \qquad \phi^A=\bar\phi^A ,
\end{equation}
We assume, as usual in this large-charge EFT analysis, that the transverse moduli are constant on the saddle. The dilaton equation fixes the norm of the angular velocity relative to the conformal mass,
\begin{equation}\label{eq:saddle-dilaton-equation}
  \mu^a \mu^b\, \widehat G_{ab}(\bar\phi) = m_d^2\,\widehat G_{\Phi\Phi}(\bar\phi),
\end{equation}
while the $\phi^A$ equations impose
\begin{equation}\label{eq:saddle-moduli-equation}
  \partial_A \left( \mu^a \mu^b\widehat G_{ab}(\phi) - m_d^2\widehat G_{\Phi\Phi}(\phi) \right)_{\bar\phi} = 0 .
\end{equation}
The charge densities conjugate to the Cartan angles are, at leading order,
\begin{equation}\label{eq:charge-density-angular-velocity}
  \rho_a = v^2\, \widehat G_{ab}(\bar\phi)\mu^b .
\end{equation}
Depending on which charges are held fixed, the equations \eqref{eq:saddle-dilaton-equation}--\eqref{eq:charge-density-angular-velocity} are used either to determine the angular velocity vector $\mu^a$, or to determine the charge vector selected by a chosen angular velocity.

To compute the Casimir energy, we expand \eqref{eq:crs-moduli-eft} to quadratic order in fluctuations around the saddle \eqref{eq:helical-saddle-general}. Specifically we write
\begin{equation}
    \Phi=v+\delta\Phi,
    \qquad
    \pi^a=\mu^a t+\delta\pi^a,
    \qquad
    \phi^A=\bar\phi^A+\delta\phi^A ,
\end{equation}
and decompose the Cartan fluctuations into the component along the helical direction and the components orthogonal to it,
\begin{equation}
  \delta\pi^a = \mu^a\,\delta\pi_0 + \delta\pi_\perp^a, \qquad \overline G_{ab}\mu^a\delta\pi_\perp^b=0,
\end{equation}
where $\overline G_{ab}=\widehat G_{ab}(\bar\phi)$. Then we find the quadratic Lagrangian \begin{equation}\label{eq:fluctuation-lagrangian-new}
\begin{split}
\mathcal{L}_{\text{fluct}}
=&
\frac{1}{2}\overline{G}_{\Phi\Phi}
\left[
(\partial \delta\Phi)^2
+
m_d^2 (\partial \delta\pi_0)^2
+
4 m_d^2 \,\delta\dot{\pi}_0 \,\delta\Phi
\right]
\\
&+
\frac{1}{2}\overline{G}_{ab}
\partial^\mu \delta\pi_\perp^a
\partial_\mu \delta\pi_\perp^b
+
\overline{G}_{aB}
\partial_\mu \delta\pi_\perp^a
\partial^\mu \delta\phi^B
+
\frac{1}{2}K_{AB}^2
\partial^\mu \delta\phi^A
\partial_\mu \delta\phi^B
\\
&+
\mu^a
\overline{G}_{ab,A}
\delta\pi_\perp^b
\delta\dot{\phi}^A
+
d_{AB}\,
\delta\dot{\phi}^A
\delta\phi^B
-
\frac{1}{2}M_{AB}^2
\delta\phi^A
\delta\phi^B .
\end{split}
\end{equation}
Bars indicate evaluation at $\phi^A=\bar\phi^A$. Here $K,M,d$ are matrices whose specific form is discussed in \cite{Cuomo:2024fuy} and we will not need.

The quadratic fluctuation problem splits into two sectors. The first is the universal charged sector (the first row of \eqref{eq:fluctuation-lagrangian-new}), which includes the radial fluctuation $\delta\Phi$ and the Goldstone fluctuation $\delta\pi_0$ along the helical direction. This sector is the same as the fixed-charge fluctuation problem of a free complex scalar and does not contribute to the Casimir energy. The second sector (the last two rows of \eqref{eq:fluctuation-lagrangian-new}) contains the remaining $n-2$ real scalar modes: the Cartan
phases orthogonal to the helical direction, denoted $\delta\pi_\perp^a$, and the transverse moduli $\delta\phi^A$. This non-universal sector is the one which controls the sign of $\alpha_0$. 

From now on we focus on the non-universal sector. Let us rewrite the bottom two rows describing this sector in a simpler matrix form. Let $x$ denote the $n-2$-vector containing both the orthogonal Cartan fluctuations
and the transverse moduli,
\begin{equation}
    x=
    \begin{pmatrix}
    \pi_\perp\\
    \phi
    \end{pmatrix}.
\end{equation}
For a scalar spherical harmonic with eigenvalue $\lambda_\ell=\ell(\ell+1)$ on the unit $S^2$, the non-universal Lagrangian is
\begin{equation}\label{eq:transverse-lagrangian-KAM}
  L_\ell = \frac12\,\dot x^{\,T}K\,\dot x +\dot x^{\,T}A\,x -\frac12\,x^{\,T}\bigl(\lambda_\ell K+M\bigr)x .
\end{equation}
Here $K^T=K$ and $M^T=M$ are symmetric, and $A^T=-A$ is antisymmetric. The matrix $M$ has support only on the non-Cartan moduli,
\begin{equation}
M=
\begin{pmatrix}
0&0\\
0&{\cal M}_{AB}
\end{pmatrix}.
\end{equation}
Since $K$ is positive definite, we can make a constant linear change of variables $x=UX$ with
\begin{equation}
  U^T K U=\mathbf 1 .
\end{equation}
Choosing $U$ block upper triangular preserves the block form of the matrices above, and preserves their (anti)symmetry. In canonical variables the Lagrangian becomes (renaming the rotated matrices $A,M$ back to $A,M$ for simplicity)
\begin{equation}\label{eq:canonical-transverse-lagrangian}
  L_\ell = \frac12\,\dot X^{\,T}\dot X +\dot X^{\,T}AX -\frac12\,X^{\,T}\bigl(\lambda_\ell \mathbf 1 + M\bigr)X.
\end{equation}

We can now discuss computing the Casimir energy for this two-derivative EFT. The normal-mode frequencies are obtained from
\begin{equation}\label{eq:general-spectrum-determinant}
  \det\Bigl((-\omega^{2}+\lambda_\ell)\mathbf 1-2i\omega A+M\Bigr)=0.
\end{equation}
The physical frequencies $\omega_{\ell,i}$ for $i=1,\ldots,n-2$ are the positive normal-mode frequencies of the quadratic Hamiltonian. The scalar contribution to $\alpha_0$ is then the Casimir energy for this Hamiltonian:
\begin{equation}
    \alpha_0^{\rm scalar}=E_{\rm Cas}
    =
    \frac12
    \sum_{\ell=0}^{\infty}
    n_\ell
    \sum_{i=1}^{n-2}
    \omega_{\ell,i}
    \qquad
    n_\ell=2\ell+1 .
    \label{eq:scalar-casimir-general}
\end{equation}
We interpret this equation as being regularized in dimensional regularization.

In general, we do not have closed-form expressions for the frequencies $\omega_{\ell,i}$. An exception is when $A$ and $M$ commute:
\begin{equation}\label{eq:A_M_commute}
  [A,M]=0,
\end{equation}
in which case one can diagonalize them simultaneously into elementary blocks. There are $1\times1$ blocks with
\begin{equation}
  A=0,\qquad M=m^2,
\end{equation}
which gives a relativistic dispersion relation with mass squared $m^2$,
\begin{equation}\label{eq:spectrum1-new}
  \omega_\ell=\sqrt{\lambda_\ell+m^2},
\end{equation}
and $2\times2$ blocks with
\begin{equation}
    A=
    \begin{pmatrix}
    0&-\nu\\
    \nu&0
    \end{pmatrix},
    \qquad
    M=m^2\,\mathbf 1_2,
\end{equation}
which give
\begin{equation}\label{eq:spectrum2-new}
  \omega_{\ell,\pm} = \left|\sqrt{\lambda_\ell+m^2+\nu^2}\ \pm\ \nu\right|.
\end{equation}
For $m^2\geq0$, the absolute values do not fold and the two branches together contribute as two relativistic dispersion relations: 
\begin{equation}
  \omega_{\ell,+}+\omega_{\ell,-} = 2\sqrt{\lambda_\ell+m^2+\nu^2}.
\end{equation}
Thus each such block reduces to ordinary relativistic scalar sums with non-negative mass squared. In the notation of \eqref{eq:xi}, each relativistic mode $\omega_\ell=\sqrt{\lambda_\ell+m^2}$ contributes $\xi(m^2-\frac14)$ to the Casimir energy. Since $\xi(m^2-\frac14)$ is non-positive for $m^2\geq0$, each such relativistic scalar contribution is non-positive. However generically the modes are non-relativistic and so the proof of non-positivity is more involved.

\subsection{Proof of the convexity bound when $M\geq 0$}\label{subsec:projected-convexity-proof}

In this section we assume that the mass matrix $M$ in \eqref{eq:canonical-transverse-lagrangian} is positive semidefinite $M\geq 0$. We show that under this assumption, the scalar contribution to $\alpha_0$ is non-positive,
\begin{equation}
  \alpha_{0}^{\rm scalar} = E_{\rm Cas} \leq 0 ,
\end{equation}

The proof follows directly from the quadratic Hamiltonian corresponding to the Lagrangian \eqref{eq:canonical-transverse-lagrangian}. The canonical momentum is
\begin{equation}
  P=\dot X+AX,
\end{equation}
so the Hamiltonian for angular momentum $\ell$ is
\begin{equation}\label{eq:general-transverse-hamiltonian}
  H_\ell = \frac12(P-AX)^T(P-AX) + \frac12 X^T\left(\lambda_\ell\mathbf 1+M\right)X .
\end{equation}
Hamilton's equations imply
\begin{equation}
  \ddot X+2A\dot X+\left(\lambda_\ell\mathbf 1+M\right)X=0,
\end{equation}
which is equivalent to the determinant condition \eqref{eq:general-spectrum-determinant}. The ground-state energy at fixed
$\ell$ is
\begin{equation}\label{eq:ell-zero-point-energy}
  E_\ell = \frac12\sum_i\omega_{\ell,i}.
\end{equation}

We now show that if $M$ is positive semidefinite $M\geq 0$, then the Casimir energy is non-positive. Introduce the reference Hamiltonian
\begin{equation}\label{eq:reference-oscillator}
  H_{\ell,\rm ref} =H_\ell+P^TAX= \frac12P^TP + \frac12 X^T\left(\lambda_\ell\mathbf 1+M-A^2\right)X .
\end{equation}
Since $-A^2=A^TA\geq0$, the matrix
$\lambda_\ell\mathbf 1+M-A^2$ is positive semidefinite. Hence the reference
ground-state energy is
\begin{equation}\label{eq:reference-energy}
  E_{\ell,\rm ref} = \frac12\operatorname{tr}\sqrt{\lambda_\ell\mathbf 1+M-A^2}.
\end{equation}
Equivalently, after diagonalizing $M-A^2$, the reference frequencies are
\begin{equation}
  \omega^{\rm ref}_{\ell,i}=\sqrt{\lambda_\ell+\kappa_i},
\end{equation}
where $\kappa_i\geq0$ are the eigenvalues of $M-A^2$.

By the variational principle, the true ground-state energy is no larger than the expectation value in any normalized trial state. We denote by $|0\rangle$ the ground state of $H_\ell$ and use the ground state $|0_{\rm ref}\rangle$ of $H_{\ell,{\rm ref}}$ as the trial state. The two Hamiltonians differ only by the mixed term:
\begin{equation}
  H_\ell = H_{\ell,\rm ref}-P^TAX.
\end{equation}
The reference ground state is a centered real Gaussian. In this state
\begin{equation}
  \langle 0_{\rm ref}|P_iX_j|0_{\rm ref}\rangle = -\frac{i}{2}\delta_{ij},
\end{equation}
and therefore
\begin{equation}
  \langle 0_{\rm ref}|P^TAX|0_{\rm ref}\rangle = -\frac{i}{2}\operatorname{tr}A =0,
\end{equation}
because $A$ is antisymmetric. Thus the variational principle tells us that
\begin{equation}
E_\ell = \langle 0|H_\ell|0 \rangle
\leq
\langle 0_{\rm ref}|H_\ell|0_{\rm ref}\rangle
=\langle 0_{\rm ref}|  H_{\ell,{\rm ref}}-P^TAX|0_{\rm ref} \rangle=
E_{\ell,\rm ref}.
\end{equation}
Here we used the fact that $M$ is positive semidefinite, since using the variational principle requires that $H_{\ell}$ is bounded from below. 

Using \eqref{eq:ell-zero-point-energy} and \eqref{eq:reference-energy}, the inequality $E_\ell\leq E_{\ell,{\rm ref}}$ gives us
\begin{equation}\label{eq:pointwise-spectral-bound}
  \sum_i\omega_{\ell,i} \leq \sum_i \sqrt{\lambda_\ell+\kappa_i}.
\end{equation}
This inequality is the main point behind the proof of the convexity bound $\alpha_0\leq 0$. The scalar contribution to $\alpha_0$ is the regulated Casimir energy
\begin{equation}
  E_{\rm Cas}=\frac12\sum_{\ell,i}(2\ell+1) \omega_{\ell,i}.
\end{equation}
We would like to compare this to the regulated Casimir energy of the reference Hamiltonian 
\begin{equation}
  E_{\rm Cas}^{\rm ref}=\frac12\sum_{\ell,i} (2\ell+1)\sqrt{\lambda_\ell+\kappa_i}.
\end{equation}
Both expressions should be understood as being regularized in dimensional regularization. Since the reference Hamiltonian has relativistic dispersion relations as in \eqref{eq:spectrum1-new} with positive $\kappa_i$, the discussion around \eqref{eq:spectrum1-new} tells us that
\begin{equation}\label{eq:reference-casimir-negative}
  E_{\rm Cas}^{\rm ref}\leq0.
\end{equation}
Then if we could use the pointwise inequality \eqref{eq:pointwise-spectral-bound} directly in the sum, this would immediately prove $E_{\rm Cas}\leq E_{\rm Cas}^{\rm ref}\leq 0$. However, since both expressions require regularization, we must be more careful. Instead it can be easily checked that the term-by-term difference between the sums is finite:
\begin{equation}
  D\equiv \frac12\sum_{\ell}(2\ell+1)\left(\sum_i\omega_{\ell,i}^{\rm ref}-\sum_i\omega_{\ell,i}\right)<\infty,
\end{equation}
and then we can use the pointwise inequality \eqref{eq:pointwise-spectral-bound} to show that $D\geq 0$. Since $D$ is finite we can write for the regulated sums
\begin{equation}
  E_{\rm Cas}^{\rm ref}-E_{\rm Cas}=D.
\end{equation}
Therefore
\begin{equation}
  E_{\rm Cas} = E_{\rm Cas}^{\rm ref}-D \leq E_{\rm Cas}^{\rm ref} \leq0.
\end{equation}
This proves that $M\geq0$ is sufficient for the scalar contribution to $\alpha_0$ to be non-positive. Together with the argument that the fermions and gauge bosons also have non-positive contributions \cite{Cuomo:2024fuy}, this proves that $\alpha_0\leq 0$ whenever the scalar matrix $M$ is positive semidefinite.

\subsection{Proof of projected convexity}\label{subsec:projected-one-charge-convexity}

We now show that for the saddle which computes the projected observable $\Dm^{(n)}(Q)$ in \eqref{eq:projected-one-charge-observable}, we must have $M\geq0$. Then the results of the previous subsection complete the proof of the bound $\alpha_0^{(n)}\leq 0$. 

Choose a single Cartan generator $T_n=n_aT^a$ and fix only the charge $Q$ under this generator. For the projected saddle, the angular velocity is along the chosen generator,
\begin{equation}\label{eq:projected-angular-velocity}
  \mu^a=\mu n^a .
\end{equation}
This does not mean that the charge vector itself is parallel to $n^a$. At the saddle value $\bar\phi$, the charge vector follows from \eqref{eq:charge-density-angular-velocity} and the constraint $n\cdot\rho=Q$:
\begin{equation}\label{eq:projected-induced-charge-vector}
  \rho_a = Q\, \frac{\widehat G_{ab}(\bar\phi)n^b} {n^c\widehat G_{cd}(\bar\phi)n^d}.
\end{equation}
Thus the state which minimizes the projected observable may carry orthogonal Cartan charges. These charges are nevertheless still of order $Q$; for any Cartan direction $f^a$,
\begin{equation}\label{eq:projected-induced-charge-bound}
  |f^a\rho_a| \leq |Q| \left[ \frac{f^a\widehat G_{ab}(\bar\phi)f^b} {n^c\widehat G_{cd}(\bar\phi)n^d} \right]^{1/2}.
\end{equation}
Allowing the orthogonal charges to vary therefore does not take us outside the large-charge regime.

We now prove the key positivity statement. The zero-mode part ($\ell=0$) of the Hamiltonian \eqref{eq:general-transverse-hamiltonian} is
\begin{equation}\label{eq:projected-zero-mode-hamiltonian}
  H_0(X,P) = \frac12(P-AX)^T(P-AX) + \frac12 X^TMX .
\end{equation}
The important point is the meaning of the canonical momentum $P$. The fields $\pi_\perp$ are the Goldstone bosons corresponding to Cartans of the broken symmetry group orthogonal to $T_n$, so their canonical momenta (which are components of $P$) are precisely the corresponding charge fluctuations. Since the projected observable fixes only the charge under $T_n$ (which sits in the ``universal sector'' we have stripped out), these orthogonal charge fluctuations are not held fixed. Therefore, if the saddle really computes $\Delta_{\min}^{(n)}(Q)$, it must be a local minimum of the unreduced Hamiltonian \eqref{eq:projected-zero-mode-hamiltonian} with respect to all small $X,P$ fluctuations. Suppose that $M$ had a negative direction. Then for some vector $v$,
\begin{equation}
  v^TMv<0 .
\end{equation}
Because the orthogonal charges are not fixed, the fluctuation
\begin{equation}\label{eq:projected-instability-fluctuation}
  X=\epsilon v, \qquad P=AX
\end{equation}
is an allowed nearby configuration at the same fixed $Q_n$. Along this
fluctuation
\begin{equation}
  H_0(\epsilon v,A\epsilon v) = \frac12\epsilon^2 v^TMv <0 .
\end{equation}
This is a nearby configuration in the same projected fixed-charge problem with lower energy. Hence the original saddle could not have been the ground-state saddle for $\Delta_{\min}^{(n)}(Q)$. We conclude that every projected one-charge ground-state saddle obeys
\begin{equation}\label{eq:projected-M-positive}
  M\geq0 .
\end{equation}

Combining \eqref{eq:projected-M-positive} with the variational theorem of the previous subsection proves that the scalar one-loop contribution to $\Delta_{\min}^{(n)}(Q)$ is non-positive, proving the convexity bound $\alpha_0^{(n)}\leq 0$.

It is useful to translate \eqref{eq:projected-M-positive} back into the EFT data, since then $\alpha_0\leq 0$ gives an explicit bound on the coefficients of the EFT. Before bringing the kinetic matrix to the identity, the non-Goldstone block of the potential in \eqref{eq:transverse-lagrangian-KAM} is ${\cal M}_{AB}$, and the sign
condition $M\geq0$ is equivalent to
\begin{equation}
  {\cal M}_{AB}\geq0.
\end{equation}
For the projected saddle \eqref{eq:projected-angular-velocity}, define the fixed-angular-velocity inertia
\begin{equation}\label{eq:projected-inertia}
  {\cal I}_n(\phi) = \frac{\mu^2 n^an^b\widehat G_{ab}(\phi)} {m_d^2\widehat G_{\Phi\Phi}(\phi)} .
\end{equation}
The saddle equations \eqref{eq:saddle-dilaton-equation} and \eqref{eq:saddle-moduli-equation} imply
\begin{equation}
  {\cal I}_n(\bar\phi)=1, \qquad \partial_A{\cal I}_n(\bar\phi)=0 .
\end{equation}
Indeed, the first equation is the dilaton equation, while the second is the $\phi^A$ equation at fixed angular velocity. At the saddle, the expression
\begin{equation}
  m_d^2\widehat G_{\Phi\Phi}(\phi) - \mu^2n^an^b\widehat G_{ab}(\phi) = m_d^2\widehat G_{\Phi\Phi}(\phi) \left(1-{\cal I}_n(\phi)\right)
\end{equation}
vanishes together with its first derivative. Hence the non-Goldstone mass matrix can be written as
\begin{equation}\label{eq:projected-M-inertia}
  {\cal M}_{AB} = -\frac12\,m_d^2\,\overline{G}_{\Phi\Phi}\, \partial_A\partial_B{\cal I}_n(\bar\phi).
\end{equation}
Thus the result ${\cal M}_{AB}\geq0$ can be translated explicitly into a condition on the EFT coefficients.

\subsection{Fixed-charge-ray convexity}\label{subsec:fixed-ray-limitations}

We now discuss convexity for $\Dm$ where we fix the full charge vector. After explaining why $M$ is not necessarily positive-semidefinite on the corresponding saddle, we discuss specific cases in which it is still possible to prove fixed-charge-ray convexity. We then describe a consistent EFT (which does not correspond to any CFT that we are aware of) which does not obey fixed-charge-ray convexity. As a result this statement, if true, cannot be proven using large-charge EFT methods alone.

\subsubsection{Why the proof fails for fixed charge vectors}\label{subsubsec:fixed-ray-variational-failure}

The argument in section \ref{subsec:projected-one-charge-convexity} does not prove fixed-charge-ray convexity. In the fully fixed charge-vector problem, the orthogonal Cartan charges are held fixed. As a result, we now cannot vary all the momenta $P$, since some of them will correspond to conserved charges which we are now holding fixed. We therefore cannot use the fluctuation \eqref{eq:projected-instability-fluctuation} to argue for an instability, since this fluctuation may change the value of the charge vector. Instead the best we can do is set $P=0$ and demand stability there, which requires
\begin{equation}\label{eq:weaker_cond}
  M-A^2\geq 0.
\end{equation}
This condition is weaker than the condition $M\geq 0$, which is the one required to prove the convexity bound. 

\subsubsection{Cases where the convexity bound can still be proven}

There are still special situations where $M\geq0$ can be proven for a fixed charge vector by other
means. First, as already discussed in \cite{Cuomo:2024fuy}, if the theory effectively has only one Cartan participating in the large-charge saddle, then there are no orthogonal Cartan charge fluctuations, and effectively $A=0$ so that the weaker condition \eqref{eq:weaker_cond} becomes $M\geq 0$. Charge conjugation or another discrete symmetry can also forbid the mixing term $\dot\pi^TA\phi$, in which case $A=0$ again.

Another useful class of models where $M\geq 0$ for the fixed charge vector case are theories with low-dimensional moduli spaces. Specifically, moduli spaces of dimension $n\leq 3$ must have $A=0$, since $A$ is an antisymmetric matrix acting on $n-2$ degrees of freedom. As a result we find that $M\geq 0$ for these low-dimensional moduli spaces.

Another useful class consists of dilaton plus NLSM EFTs where the sigma-model target is locally symmetric,
\begin{equation}\label{eq:locally-symmetric-condition-main}
  \nabla_E R^A{}_{BCD}=0 .
\end{equation}
In appendix~\ref{app:locally-symmetric} we show that for this class the matrices $A$ and $M$ defined in \eqref{eq:canonical-transverse-lagrangian} commute. As discussed around \eqref{eq:A_M_commute}, the result is that the spectrum includes only relativistic dispersion relations (up to shifts of $\pm \nu$ for some $\nu$ which cancel in pairs when we sum over the frequencies). This almost proves that the Casimir energy is non-positive, except that we first must prove that the mass matrix $M$ only has positive eigenvalues, i.e.~that $M\geq 0$. This must be checked explicitly for specific examples, and indeed for the relevant cases in this paper (the $\mathbb{CP}^{n}$ model in section \ref{sec:SQED} and the $\frac{Sp(N_f)}{Sp(N_f-1)\times Sp(1)}$ target space in section \ref{sec:SU2_SQCD}) we indeed find $M\geq 0$. For compact \textit{maximally} symmetric targets this sign can be checked universally, and one finds $M\geq0$ in general, so the fixed-charge-ray convexity bound follows in this subclass.

\subsubsection{A two-derivative EFT counterexample}\label{subsubsec:eft-counterexample-every-ray}

We now give a simple EFT example showing why fixed-charge-ray convexity cannot be derived from the two-derivative moduli-space EFT alone. Consider the 3d EFT
\begin{equation}
\mathcal L=\frac12(\partial\Phi)^2-\frac{1}{8}\Phi^2+\frac c2\Phi^2 h_{ij}(x)\,
\partial_\mu\varphi^i\partial^\mu\varphi^j,
\qquad
c=\frac14,
\end{equation}
with target coordinates
\begin{equation}
  \varphi^i=(\theta_1,\theta_2,x)
\end{equation}
and compact target metric
\begin{equation}\label{eq:counterexample-target-metric}
  ds_h^2 = dx^2 + \bigl(1+f(x)^2\bigr)d\theta_1^2 + 4f(x)d\theta_1d\theta_2 + d\theta_2^2, \qquad f(x)=\frac12\sin(2x).
\end{equation}
The torus block is
\begin{equation}
h_{ab}
=
\begin{pmatrix}
1+f^2 & 2f\\
2f & 1
\end{pmatrix},
\qquad
\det h_{ab}=1-3f^2\geq\frac14,
\end{equation}
so the metric is everywhere positive.

There are two charges corresponding to translations in $\theta_1,\theta_2$. First fix the full charge vector
\begin{equation}
  \rho_a=(\rho,0).
\end{equation}
The homogeneous fixed-charge energy is
\begin{equation}\label{eq:counterexample-fixed-vector-energy}
  \mathcal H_Q(x) = \frac{|\rho|}{2\sqrt c} \sqrt{h^{11}(x)} = \frac{|\rho|}{2\sqrt c} \frac{1}{\sqrt{1-3f(x)^2}}.
\end{equation}
Therefore $x=0$ is a global minimum in this fully fixed charge-vector sector. The corresponding helical saddle is
\begin{equation}\label{eq:counterexample-fixed-vector-saddle}
  \Phi=v, \qquad \theta_1=t, \qquad \theta_2=0, \qquad x=0.
\end{equation}
It obeys the dilaton equation because
\begin{equation}
  c\,\dot\theta_1^{\,2} = \frac{1}{4},
\end{equation}
and the $x$-equation because
\begin{equation}
    \partial_x h_{11}\,\dot\theta_1^{\,2}
    =
    2ff'\,\dot\theta_1^{\,2}
    =
    0
    \qquad
    \text{at }x=0.
\end{equation}

Expanding the non-universal sector $(\theta_2,x)$ around the saddle gives, after canonical normalization,
\begin{equation}\label{eq:counterexample-AM}
    A
    =
    \begin{pmatrix}
    0&-1\\
    1&0
    \end{pmatrix},
    \qquad
    M
    =
    \begin{pmatrix}
    0&0\\
    0&-1
    \end{pmatrix}.
\end{equation}
Thus the weaker condition $M-A^2\geq 0$ is met, but we have $M\not\geq0$. 

The normal-mode frequencies obey
\begin{equation}\label{eq:counterexample-boson-frequencies}
  \omega_{\ell,\pm}^2 = \frac{ 2\lambda_\ell+3 \pm \sqrt{16\lambda_\ell+9} }{2}, \qquad \lambda_\ell=\ell(\ell+1).
\end{equation}
The lower branch is non-negative for all $\ell$, with a zero only at $\ell=0$. The regulated scalar Casimir energy in dimensional regularization is
\begin{equation}\label{eq:counterexample-boson-casimir}
E_{\rm Cas}^{\rm scalar}
=
\sum_{\ell=0}^{\infty}
\left[
r_\ell
\left(
\omega_{\ell,+}+\omega_{\ell,-}
\right)
-
\left(
2r_\ell^2+\frac14
\right)
\right],
\qquad
r_\ell=\ell+\frac12 .
\end{equation}
Numerically, we find
\begin{equation}
  E_{\rm Cas}^{\rm scalar} \simeq 0.02868 > 0 .
\end{equation}

To make the example a counterexample to the convexity bound in a 3d $\mathcal N=1$ moduli-space EFT, we must also include the contribution from the fermions. In an orthonormal frame on the full target-space cone, let $Y^I=(\Phi,\theta_1,\theta_2,x)$. Hatted indices label tangent-frame directions, $\chi_{\hat I}$ are the Majorana fermions paired with the bosonic moduli, $\gamma^\mu$ are the cylinder gamma matrices, $\nabla_\mu$ is the spacetime spinor covariant derivative, and $\omega_{K,\hat I\hat J}$ is the target-space spin connection. The quadratic fermion action around a bosonic background is
\begin{equation}
    \mathcal L_{\rm f}^{(2)}
    =
    \frac{i}{2}
    \bar\chi_{\hat I}\gamma^\mu
    \left(
    \nabla_\mu\delta_{\hat I\hat J}
    +
    \partial_\mu Y^K\omega_{K,\hat I\hat J}
    \right)
    \chi_{\hat J}.
\end{equation}
On the saddle, the only nonzero background derivative is $\dot\theta_1=1$. At $x=0$, the metric $h$ is locally the identity in the coordinates $(\theta_1,\theta_2,x)$, while
\begin{equation}
  h_{12}=2f(x), \qquad \partial_x h_{12}\big|_{x=0}=2 .
\end{equation}
Thus the base spin connection has
\begin{equation}
  \bar\omega_{\theta_1,\hat\theta_2\hat x}=1
\end{equation}
at the saddle. The cone factor gives the additional universal component
\begin{equation}
  \omega_{\theta_1,\hat\Phi\hat\theta_1}=\sqrt c .
\end{equation}
The pulled-back target spin connection therefore splits the four Majorana fermions into two blocks:
\begin{equation}
    \Omega_{\hat\Phi\hat\theta_1}
    =
    \sqrt c\,\dot\theta_1
    =
    \frac{1}{2},
    \qquad
    \Omega_{\hat\theta_2\hat x}
    =
    \dot\theta_1
    =
    1 .
\end{equation}
The first block is the universal radial/longitudinal pair, and the second is the non-universal pair associated with $(\theta_2,x)$.

For a two-Majorana block with dimensionless chemical potential $\nu$, the spinor frequencies on $S^2$ are
\begin{equation}
  \omega_{n,\pm} = n\pm\nu \qquad n=1,2,\ldots,
\end{equation}
with degeneracy $2n$. In dimensional regularization the contribution is
\begin{equation}
  E_{\rm f}(\nu) = -2 \sum_{n=1}^{\infty}n(\nu-n)_+ .
\end{equation}
Here the two blocks have
\begin{equation}
  \nu_{\rm univ}=\frac12, \qquad \nu_{\perp}=1.
\end{equation}
Therefore neither block has a negative-energy filled level. The $\nu_\perp=1$ block has a zero mode at the first spinor level, but it does not shift the Casimir energy. Thus
\begin{equation}
  E_{\rm Cas}^{\rm ferm}=0,
\end{equation}
and the total one-loop contribution remains
\begin{equation}\label{eq:counterexample-total-casimir}
    E_{\rm Cas}^{\rm total}
    =
    E_{\rm Cas}^{\rm scalar}
    +
    E_{\rm Cas}^{\rm ferm}
    \simeq
    0.02868
    >
    0 .
\end{equation}
This example is thus an EFT counterexample to the statement that every fixed ray in charge space has non-positive $\alpha_0$.

Let us finally compare this with the projected observable. If we fix only the $\theta_1$ charge, then at fixed $x$ the minimizing charge vector is given
by \eqref{eq:projected-induced-charge-vector} with $n^a=(1,0)$:
\begin{equation}
  \rho_1=Q, \qquad \rho_2 = Q\frac{h_{21}}{h_{11}} = Q\frac{2f}{1+f^2}.
\end{equation}
The projected fixed-$Q_1$ energy is proportional to $(h_{11})^{-1/2}$, so it is minimized by maximizing $h_{11}=1+f^2$. Since $|f|\leq 1/2$, the projected saddle lies at $f=\pm1/2$, not at $f=0$. At this projected saddle,
\begin{equation}
  Q_2=\rho_2=\pm\frac45 Q_1 ,
\end{equation}
and we find
\begin{equation}
    A_{\rm proj}=0,
    \qquad
    M_{\rm proj}
    =
    \begin{pmatrix}
    0&0\\
    0&\frac45
    \end{pmatrix}.
\end{equation}
Thus $M_{\rm proj}\geq0$, as expected from the argument above for the projected saddle.

\section{Bounds on $\alpha_1$}\label{sec:alpha1}

We now turn to the leading coefficient $\alpha_1$, which measures the slope of the tower. An immediate trivial bound following from the standard unitarity bound is
\begin{equation}
  \alpha_1\geq0.
\end{equation}
The question is whether there is any stronger universal bound.

The main motivation for such a bound comes from holography and scale separation. In a simple bulk picture, a linear tower of charged operators can arise from Kaluza--Klein momentum on an internal circle. If the bulk contains
an $AdS_d\times S^1_R$ region, with AdS radius $L$, then a KK mode with integer momentum $n$ has
\begin{equation}
  m_n^2=m_0^2+\frac{n^2}{R^2}.
\end{equation}
At large $|n|$, the dual operator dimension behaves as
\begin{equation}
  \Delta(n)\sim Lm_n\sim \frac{L}{R}|n|,
\end{equation}
so in this cartoon one would identify
\begin{equation}
  \alpha_1\sim \frac{L}{R}.
\end{equation}
A theory with parametrically small internal radius would therefore have a parametrically large value of $\alpha_1$. Conversely, an absence of parametric scale separation might suggest that $\alpha_1$ should remain order one. 

This interpretation is most natural when the charge is associated with a geometric isometry whose normalization is fixed, for example a Cartan of a simple non-Abelian symmetry. In the examples of section \ref{sec:computations},
the corresponding slopes are indeed order one, and we will not find any obstruction to this expectation in such cases. The difficulties arise for Abelian symmetries. For a standalone $U(1)$, the large-charge slope is very
sensitive to the charge lattice, to finite quotients, and to the choice of a linear combination of several $U(1)$ factors. These are perfectly ordinary CFT operations, but they obscure any direct relation between the raw coefficient $\alpha_1$ and a geometric radius. In a semiclassical holographic theory, a bulk observer would also keep track of the full charge lattice, the current two-point matrix, and multi-particle states. Thus the arguments below should not be interpreted as producing genuine bulk scale separation in a large-$N$ theory. Rather, they show that $\alpha_1$ by itself is not the right CFT observable for diagnosing such scale separation.

Let us see how this happens. Before asking whether $\alpha_1$ is bounded, one must specify how the charge
is normalized. Throughout this discussion we use the natural CFT normalization: the charge $Q$ is normalized so that the smallest nonzero charge carried by a local operator is one. With this convention,
\begin{equation}\label{eq:tower_temp}
  \Delta_{\min}(Q)=\alpha_1 Q+\cdots
\end{equation}
defines the coefficient $\alpha_1$. There are now simple operations on a CFT which can change $\alpha_1$ by an arbitrarily large or arbitrarily small factor:
\begin{itemize}
    \item First, $\alpha_1$ has no universal upper bound. Suppose a CFT has a $U(1)$
global symmetry and a large-charge tower as in \eqref{eq:tower_temp}. Now gauge a finite subgroup
\begin{equation}
  \mathbb Z_N\subset U(1).
\end{equation}
This gives a new fully consistent CFT. The continuous global symmetry of the new theory is $U(1)/\mathbb Z_N$, and local operators whose original charge was not divisible by $N$ are projected out. Therefore the surviving large-charge tower is obtained from the original one by setting
\begin{equation}
  Q_{\rm old}=N Q_{\rm new},
\end{equation}
where $Q_{\rm new}$ is again normalized so that the smallest surviving charge
is one. Thus
\begin{equation}
  \Delta_{\min}^{\rm new}(Q_{\rm new}) = \alpha_1\,N Q_{\rm new}+\cdots,
\end{equation}
and so
\begin{equation}
  \alpha_1^{\rm new}=N\alpha_1.
\end{equation}
Since $N$ can be chosen arbitrarily large, there is no universal upper bound on $\alpha_1$ itself.
\item There is also no universal lower bound. This time the simplest mechanism is to choose an unusual primitive direction in a multi-$U(1)$ charge lattice. Suppose the theory has two commuting charges $Q_x,Q_y$, and that the leading large-charge answer in this plane is
\begin{equation}
  \Delta_{\min}(Q_x,Q_y) = \frac{1}{\sqrt2}\sqrt{Q_x^2+Q_y^2}+\cdots . \label{eq:alpha1-two-charge-example}
\end{equation}
This is precisely the leading behavior found in the large-$N$ $SO(N)\times SO(N)$ example of section \ref{sec:SO(N)SO(N)_epsilon}. Now choose a primitive $U(1)$ generator
\begin{equation}
  Q=aQ_x+bQ_y, \qquad \gcd(a,b)=1.
\end{equation}
The condition $\gcd(a,b)=1$ means that the charge $Q$ is already normalized so that the charge lattice contains states of charge one.

If we only fix this single charge $Q$, the lightest state at large $Q$ is obtained by minimizing \eqref{eq:alpha1-two-charge-example} over $Q_x,Q_y$ subject to
\begin{equation}
  aQ_x+bQ_y=Q.
\end{equation}
Treating the charges as continuous is sufficient at leading order in large $Q$, since lattice effects only change the answer by $O(Q^0)$. The minimum is obtained for a charge vector parallel to $(a,b)$,
\begin{equation}
  (Q_x,Q_y) = \frac{Q}{a^2+b^2}(a,b)+O(Q^0),
\end{equation}
and therefore
\begin{equation}
  \Delta_{\min}^{(a,b)}(Q) = \frac{1}{\sqrt2}\, \frac{Q}{\sqrt{a^2+b^2}} +O(Q^0).
\end{equation}
Thus
\begin{equation}
  \alpha_1^{(a,b)} = \frac{1}{\sqrt{2(a^2+b^2)}}.
\end{equation}
By taking $a^2+b^2$ large, this coefficient can be made arbitrarily small.

This example illustrates part of the problem with interpreting $\alpha_1$ as a measure of scale separation. The small slope is not telling us that an internal circle has become large. It is telling us that the chosen $U(1)$ is a long primitive vector in a higher-rank charge lattice, so that states carrying large charge under this $U(1)$ can be built from comparatively modest charges under the original Cartans.\footnote{We thank
Miguel Montero for discussions on this point.}
\end{itemize}

The conclusion is that $\alpha_1$ by itself is not a universal CFT observable with a meaningful upper or lower bound. It depends too strongly on discrete choices in the charge lattice and on which $U(1)$ direction one decides to study.

One can instead attempt to study $\alpha_1$ with a different normalization of the charge. It is tempting to suggest the following alternative. Let
\begin{equation}
  \langle J_\mu(x)J_\nu(0)\rangle = C_J\,\frac{I_{\mu\nu}(x)}{x^{2d-2}},
\end{equation}
and
\begin{equation}
  \langle T_{\mu\nu}(x)T_{\rho\sigma}(0)\rangle = C_T\,\frac{\mathcal I_{\mu\nu,\rho\sigma}(x)}{x^{2d}} ,
\end{equation}
where $J_\mu$ is the current for the $U(1)$ whose charge is being fixed, and $T_{\mu\nu}$ is the energy-momentum tensor. Here $I_{\mu\nu}$ and $I_{\mu\nu,\rho\sigma}$ are tensor structures fixed by symmetries whose specific form we will not need. Naively, $C_T$ is a measure of the number of degrees of freedom in the system, while $C_J$ is a measure of the number of charged degrees of freedom. A natural combination is then\footnote{Using $C_J$ alone would remove the charge-normalization issues discussed above, but it would make $\hat\alpha_1$ arbitrarily small (of order $1/N$) for any large-$N$ theory, making it an uninteresting observable in holographic theories.}
\begin{equation}
  \widehat\alpha_1 = \alpha_1\sqrt{\frac{C_J}{C_T}} .
\end{equation}

This normalized slope fixes the two simplest problems discussed above. Specifically, gauging a finite subgroup $\mathbb Z_N\subset U(1)$ leads to $C_J\to C_J/N^2$,
and the combination $\alpha_1\sqrt{C_J}$ is unchanged. Similarly, if we choose
a primitive direction
\begin{equation}
  Q=aQ_x+bQ_y,
\end{equation}
then the current is
\begin{equation}
  J=aJ_x+bJ_y.
\end{equation}
In an orthogonal current basis, $C_J$ grows like $a^2+b^2$, compensating the
factor $1/\sqrt{a^2+b^2}$ in \eqref{eq:alpha1-two-charge-example}.

We can still make $\hat\alpha_1$ arbitrarily small, simply by finding examples with many uncharged degrees of freedom, so that $C_T\gg C_J$. However, it seems much more difficult to find examples where $\hat\alpha_1$ is arbitrarily large, and indeed we do not yet have such an example. So it is possible that the correct upper bound is on $\hat\alpha_1$.

The downside of working with $\hat\alpha_1$ is that this same normalization also removes the naive relation to scale-separation. For a KK photon on an internal circle of radius $R$, the discussion above gave
\begin{equation}
  \alpha_1\sim \frac{L}{R}.
\end{equation}
On the other hand, dimensional reduction of the Einstein term gives
\begin{equation}
  \frac{C_J}{C_T} \sim \frac{R^2}{L^2},
\end{equation}
because the KK gauge kinetic term carries the extra factor of $R^2$. Hence
\begin{equation}
  \widehat\alpha_1 = \alpha_1\sqrt{\frac{C_J}{C_T}} \sim O(1).
\end{equation}
Thus $\widehat\alpha_1$ is better behaved as a CFT normalization, but it is not a direct measure of the hierarchy $L/R$.

We therefore do not find a universal swampland-type bound either on $\alpha_1$ itself or on the normalized slope $\widehat\alpha_1$. The raw coefficient $\alpha_1$ is the quantity which most directly resembles a KK slope, but it is too sensitive to the choice of charge lattice and to projections of higher-rank symmetries. The normalized coefficient is better behaved as a CFT quantity, but no longer directly measures the scale-separation parameter. A sharper holographic diagnostic would presumably need to keep track of the full current two-point matrix, the full charge lattice, and, in large-$N$ theories, multi-particle states. We leave this
possibility for future work.

\section*{Acknowledgments}

The authors would like to thank I. Valenzuela for many useful discussions and for the observation which sparked this project. We would also like to thank G. Cuomo, S. Hellerman and M. Watanabe for useful discussions, and J. Kulp and M. Montero for useful discussions and comments on a draft. A.S. is supported by the Mani L. Bhaumik Institute for Theoretical Physics.

\newpage

\begin{appendix}

\section{Locally symmetric target spaces and commutativity}\label{app:locally-symmetric}

In this appendix we prove that for a dilaton coupled to a NLSM, a locally symmetric target implies $[A,M]=0$ using the notation of \eqref{eq:canonical-transverse-lagrangian}. However it is not enough to prove $M\geq0$. Our dilaton-coupled-to-NLSM EFT is explicitly given by
\begin{equation}
\mathcal L
=
\frac12\,\partial_\mu\Phi\,\partial^\mu\Phi
-\frac12\,m_d^2\,\Phi^2
+\frac c2\,\Phi^2\,G_{AB}(\varphi)\,\partial_\mu\varphi^A\partial^\mu\varphi^B,
\qquad
m_d^2=\frac{(d-2)^2}{4}.
\end{equation}
We expand around a homogeneous helical saddle
\begin{equation}
  \Phi=v, \qquad \partial_i\varphi^A=0, \qquad \dot\varphi^A=\mu\,k^A(\varphi_0),
\end{equation}
where $k^A$ is a Killing vector of the target-space metric $G_{AB}$.
The equations of motion imply
\begin{equation}
  k^B\nabla_B k^A=0, \qquad m_d^2=c\,\mu^2 k^2.
\end{equation}
Thus the orbit generated by $k^A$ is a geodesic.
Let $\eta^A$ be the covariant target-space fluctuation and decompose it into a longitudinal mode and transverse modes,
\begin{equation}
\eta^A=\eta_\parallel\,\hat k^A+\eta_\perp^A,
\qquad
\hat k^A=\frac{k^A}{|k|},
\qquad
k_A\eta_\perp^A=0.
\end{equation}
The dilaton mixes only with the longitudinal mode $\eta_\parallel$, giving the universal sector discussed in section \ref{subsec:alpha0-setup}. The non-universal contribution to the Casimir energy comes entirely from the transverse fluctuations $\eta_\perp^A$. After expanding them in spherical harmonics on $S^{d-1}$, the quadratic Lagrangian for each angular momentum $\ell$ takes the form
\begin{equation}
  L_\ell := \frac12\,\dot q^{\,T}\dot q +\dot q^{\,T}A\,q -\frac12\,q^{\,T}\bigl(\lambda_\ell \mathbf 1 + M\bigr)q,
\end{equation}
with $q$ the vector of transverse modes. In the present case the matrices entering this quadratic problem are
\begin{equation}
A^A{}_B=\mu\,\nabla_B k^A\big|_{\perp},
\qquad
M^A{}_B=B^A{}_B+(A^2)^A{}_B,
\qquad
B^A{}_B=\mu^2 R^A{}_{CBD}\,k^C k^D\big|_{\perp}.
\label{eq:appendix-ABM}
\end{equation}

If the target space is locally symmetric,
\begin{equation}
  \nabla_E R^A{}_{BCD}=0.
\end{equation}
Using also the geodesic condition $k^B\nabla_B k^A=0$, we find
\begin{equation}
  k^E\nabla_E B^A{}_B = \mu^2 k^E\nabla_E\left(R^A{}_{CBD}k^Ck^D\right) =0.
\end{equation}
Since $k^A$ is Killing, its flow preserves the curvature tensor and therefore also preserves $B$:
\begin{equation}
  \mathcal L_k B^A{}_B=0.
\end{equation}
Using the Lie derivative of a $(1,1)$ tensor,
\begin{equation}
  (\mathcal L_k B)^A{}_B = k^E\nabla_E B^A{}_B -B^E{}_B\,\nabla_E k^A +B^A{}_E\,\nabla_B k^E,
\end{equation}
and $A^A{}_B=\mu\,\nabla_B k^A$, we conclude that
\begin{equation}
  [A,B]=0.
\end{equation}
Since $M=B+A^2$, this immediately implies
\begin{equation}\label{eq:locally-symmetric-AM-commute}
  [A,M]=0.
\end{equation}

As discussed around \eqref{eq:A_M_commute}, the commutativity statement is not yet enough to prove convexity: we still need the sign condition $M\geq0$. We can prove this extra condition under the stronger assumption that
the target space is maximally symmetric. Locally, such a target is a round sphere, which we embed as
\begin{equation}
  X^I X^I=L^2 .
\end{equation}
The Killing vector which generates the helical motion can then be written as a linear rotation,
\begin{equation}
  k^I=\Omega^I{}_J X^J, \qquad \Omega^T=-\Omega .
\end{equation}
It is useful to define
\begin{equation}
  S=-\Omega^2, \qquad |k|^2=X^T S X .
\end{equation}
The fixed-charge ground state chooses the point $X_0$ on the sphere where $|k|^2$ is largest. Therefore $X_0$ is an eigenvector of $S$ with the largest eigenvalue,
\begin{equation}
  S X_0=s_* X_0, \qquad s_*=\frac{|k|^2_{X=X_0}}{L^2}=\lambda_{\max}(S).
\end{equation}
Now take any small fluctuation $Y$ tangent to the sphere at $X_0$, so $Y\cdot X_0=0$. Since $s_*$ is the largest eigenvalue of $S$, we have
\begin{equation}
  Y^T S Y\leq s_*\,Y^T Y .
\end{equation}
For the maximally symmetric target, the quadratic form defined by the matrix
$M$ is
\begin{equation}
  Y^T M Y = \mu^2\left(s_*\,Y^T Y-Y^T S Y\right),
\end{equation}
up to the positive normalization factor removed when bringing the kinetic term to canonical form. The inequality above therefore implies
\begin{equation}
  Y^T M Y\geq0
\end{equation}
for every tangent fluctuation $Y$. Thus $M\geq0$, so the missing sign condition holds for compact maximally symmetric targets.

\section{Equal-scaling regime of the $SO(N)\times SO(N)$ model}\label{app:equal-scaling}

In Section~\ref{sec:SO(N)SO(N)_epsilon} we analyzed the large-$N$ effective action~\eqref{eq:son-Seff} of the $SO(N)\times SO(N)$ model in the regime $Q_{x,y}\gg N\gg 1$, where the classical condensate terms dominate over the $\Tr\log$ determinants. There the determinants did not back-react on the leading saddle~\eqref{eq:son-saddle}, and they contributed only the subleading one-loop term in~\eqref{eq:son-Delta-final}. In this appendix we treat the complementary, ``equal-scaling'' regime in which the charges grow linearly with $N$,
\begin{equation}
  Q_{x,y} = \lambda_{x,y}\,N, \qquad N\to\infty,\quad \lambda_{x,y}\ \text{fixed}, \label{eq:app-equal-scaling-def}
\end{equation}
where we have introduced the fixed charge ratios $\lambda_{x,y}\equiv Q_{x,y}/N$. In the regime \eqref{eq:app-equal-scaling-def} the condensate and determinant contributions are both of order $N$ and must be treated on an equal footing.

We start from the large-$N$ effective action \eqref{eq:son-Seff}, repeated here for convenience,
\begin{align}
S_{\rm eff}
&=
\frac{N}{2}\Tr\log\left[-\Delta+F-a^2+\tfrac14\right]
+\frac{N}{2}\Tr\log\left[-\Delta-F-a^2+\tfrac14\right]
\notag\\
&\quad
-N\Tr\log\left[i\gamma^\mu\partial_\mu+a\right]
-N\Tr\log\left[i\gamma^\mu\partial_\mu-a\right]
\notag\\
&\quad
+i\theta_x Q_x+i\theta_y Q_y
+\left(F+\theta_x^2+\tfrac14\right)X^2
+\left(-F+\theta_y^2+\tfrac14\right)Y^2
+2a(XF_x-YF_y)+F_x^2+F_y^2,
\label{eq:app-equal-Seff}
\end{align}
We now impose the scaling~\eqref{eq:app-equal-scaling-def} and rescale the condensates as
\begin{equation}
  X=\sqrt{N}\,\bar{X}, \qquad Y=\sqrt{N}\,\bar{Y}, \label{eq:app-equal-rescale}
\end{equation}
holding $F,a,\theta_x,\theta_y,F_x,F_y=O(1)$. With \eqref{eq:app-equal-rescale} the last line of \eqref{eq:app-equal-Seff} is subleading after dividing by $N$, since
\begin{equation}
  \frac{2a(XF_x-YF_y)}{N}=O(N^{-1/2}),\qquad \frac{F_x^2+F_y^2}{N}=O(N^{-1}). \label{eq:app-equal-subleading}
\end{equation}
Using $Q_{x,y}=\lambda_{x,y}N$, the leading action per unit $N$ is therefore
\begin{align}
\frac{S_{\rm eff}}{N}
&=
\frac12\Tr\log\left[-\Delta+F-a^2+\tfrac14\right]
+\frac12\Tr\log\left[-\Delta-F-a^2+\tfrac14\right]
\notag\\
&\quad
-\Tr\log\left[i\gamma^\mu\partial_\mu+a\right]
-\Tr\log\left[i\gamma^\mu\partial_\mu-a\right]
\notag\\
&\quad
+i\theta_x \lambda_x+i\theta_y \lambda_y
+\left(F+\theta_x^2+\tfrac14\right)\bar{X}^2
+\left(-F+\theta_y^2+\tfrac14\right)\bar{Y}^2.
\label{eq:app-equal-leading}
\end{align}

Varying \eqref{eq:app-equal-leading} with respect to $\bar{X}$ and $\bar{Y}$ gives the mass conditions
\begin{align}
\left(F+\theta_x^2+\tfrac14\right)\bar{X}&=0,
\label{eq:app-equal-x-mass}\\
\left(-F+\theta_y^2+\tfrac14\right)\bar{Y}&=0,
\label{eq:app-equal-y-mass}
\end{align}
while varying with respect to $\theta_x$ and $\theta_y$ gives the charge conditions
\begin{align}
i \lambda_x+2\theta_x\bar{X}^2&=0,
\label{eq:app-equal-theta-x}\\
i \lambda_y+2\theta_y\bar{Y}^2&=0.
\label{eq:app-equal-theta-y}
\end{align}
These are the per-$N$ analogues of the fixed-charge conditions in \eqref{eq:son-saddle}. For $\lambda_{x,y}\neq 0$, equations \eqref{eq:app-equal-theta-x}--\eqref{eq:app-equal-theta-y} force $\bar{X},\bar{Y}\neq 0$, so the brackets in \eqref{eq:app-equal-x-mass}--\eqref{eq:app-equal-y-mass} must vanish identically,
\begin{equation}
  F+\theta_x^2+\tfrac14=0, \qquad -F+\theta_y^2+\tfrac14=0. \label{eq:app-equal-mass-zero}
\end{equation}
The saddle thus lies on the imaginary-$\theta$ contour, as in section~\ref{sec:SO(N)SO(N)_epsilon}. For $\lambda_{x,y}>0$ the branch compatible with \eqref{eq:app-equal-theta-x}--\eqref{eq:app-equal-theta-y} is
\begin{align}
i\theta_x&=\sqrt{F+\tfrac14},
\label{eq:app-equal-theta-x-sol}\\
i\theta_y&=\sqrt{\tfrac14-F},
\label{eq:app-equal-theta-y-sol}
\end{align}
which requires $-\tfrac14\leq F\leq\tfrac14$. Substituting \eqref{eq:app-equal-theta-x-sol}--\eqref{eq:app-equal-theta-y-sol} back into \eqref{eq:app-equal-theta-x}--\eqref{eq:app-equal-theta-y} fixes the condensates,
\begin{align}
\bar{X}^2&=\frac{\lambda_x}{2\sqrt{F+\frac14}},
\label{eq:app-equal-xbar}\\
\bar{Y}^2&=\frac{\lambda_y}{2\sqrt{\frac14-F}}.
\label{eq:app-equal-ybar}
\end{align}
Finally, the variation of \eqref{eq:app-equal-leading} with respect to $a$ is proportional to $a$ from the bosonic determinants, while the two fermionic determinant variations cancel at $a=0$. Hence, exactly as below \eqref{eq:son-saddle}, the $a=0$ branch is stationary, and we set $a=0$ from now on.

At $a=0$ the bosonic determinants in \eqref{eq:app-equal-leading} are expressed through the function $\xi$ defined in \eqref{eq:xi}, $\xi(x)\equiv\frac1T\Tr\log[-\Delta+x+\frac14]$. Their combined contribution to $S_{\rm eff}/N$ is
\begin{equation}
  \frac12\bigl[\xi(F)+\xi(-F)\bigr]
  =\frac12\sum_{l=0}^{\infty}\left(l+\tfrac12\right)
  \left[\sqrt{\left(l+\tfrac12\right)^2+F}+\sqrt{\left(l+\tfrac12\right)^2-F}-2\left(l+\tfrac12\right)\right],
\label{eq:app-equal-det-sum}
\end{equation}
where the explicit sum follows from \eqref{eq:xi} (the linear-in-$F$ counterterms cancel between $\xi(F)$ and $\xi(-F)$). The fermionic determinants are independent of $F$ and contribute only an additive constant, which we drop. On the saddle \eqref{eq:app-equal-mass-zero} the mass terms in \eqref{eq:app-equal-leading} vanish, and the charge terms reduce, using \eqref{eq:app-equal-theta-x-sol}--\eqref{eq:app-equal-theta-y-sol}, to $i\theta_x \lambda_x=\lambda_x\sqrt{F+\frac14}$ and $i\theta_y \lambda_y=\lambda_y\sqrt{\frac14-F}$. The action per $N$ on this branch is therefore
\begin{align}
\left.\frac{S_{\rm eff}}{N}\right|_{\rm branch}
&=
\frac12\bigl[\xi(F)+\xi(-F)\bigr]
+\lambda_x\sqrt{F+\tfrac14}+\lambda_y\sqrt{\tfrac14-F}.
\label{eq:app-equal-reduced}
\end{align}
The physical dimension is $\Dm/N = \left.S_{\rm eff}/N\right|_{\rm branch}$ evaluated at the saddle value $F_*$ determined below. Note that the determinant piece $\frac12[\xi(F)+\xi(-F)]$ is precisely the per-$N$ version of the one-loop term $\frac{N}{2}\xi(\cdot)+\frac{N}{2}\xi(\cdot)$ of \eqref{eq:son-Delta-final}; the difference is that here $F$ is fixed by the back-reacted saddle rather than by the leading classical value in \eqref{eq:son-saddle}.

Varying \eqref{eq:app-equal-leading} with respect to $F$ before substituting $\bar{X},\bar{Y}$ gives $\frac12[\xi'(F)-\xi'(-F)]+\bar X^2-\bar Y^2=0$. Using \eqref{eq:app-equal-xbar}--\eqref{eq:app-equal-ybar} and the derivative of \eqref{eq:xi},
\begin{equation}
  \xi'(x)=\sum_{l=0}^{\infty}\left[\frac{l+\frac12}{2\sqrt{\left(l+\frac12\right)^2+x}}-\frac12\right], \label{eq:app-equal-xi-prime}
\end{equation}
the stationarity condition becomes the single equation that determines $F_*$,
\begin{align}
0
&=
\frac14\sum_{l=0}^{\infty}
\left(l+\tfrac12\right)\left[\frac{1}{\sqrt{\left(l+\frac12\right)^2+F}}-\frac{1}{\sqrt{\left(l+\frac12\right)^2-F}}\right]
+\frac{\lambda_x}{2\sqrt{F+\frac14}}-\frac{\lambda_y}{2\sqrt{\frac14-F}}.
\label{eq:app-equal-F-eq}
\end{align}
For $\lambda_x=\lambda_y$ the right-hand side of \eqref{eq:app-equal-F-eq} is manifestly antisymmetric under $F\to -F$, so $F_*=0$ on the diagonal.

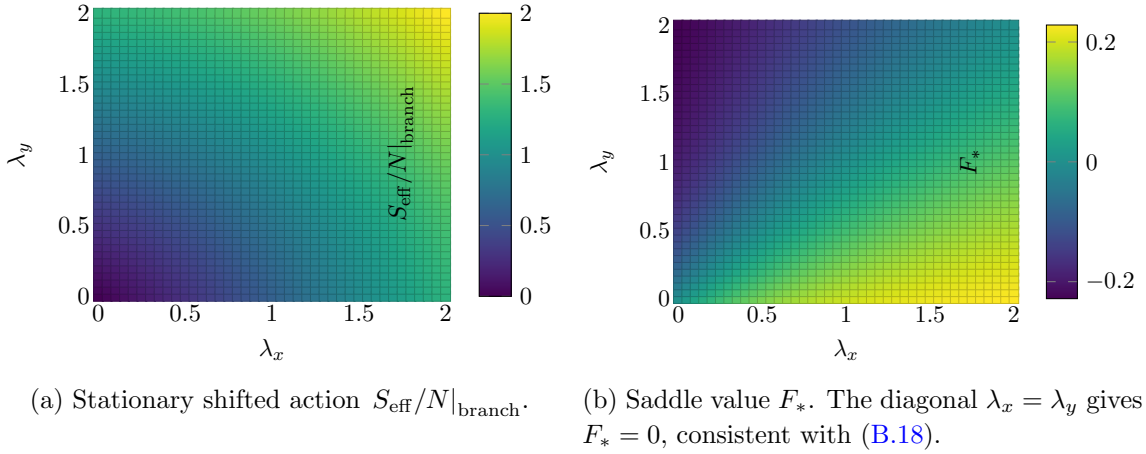
\begin{figure}[htbp]
\centering
\begin{subfigure}[t]{0.49\textwidth}
\centering
\resizebox{\linewidth}{!}{%
\begin{tikzpicture}
\begin{axis}[
    width=7cm,
    height=6cm,
    xlabel={$\lambda_x$},
    ylabel={$\lambda_y$},
    xmin=0, xmax=2,
    ymin=0, ymax=2,
    colorbar,
    colorbar style={ylabel={$\left.S_{\rm eff}/N\right|_{\rm branch}$}},
    colormap/viridis,
]
\addplot[
    scatter,
    only marks,
    mark=square*,
    mark size=2.2pt,
    scatter src=explicit,
] table[
    col sep=comma,
    x=alpha_x,
    y=alpha_y,
    meta=s_eff_per_N_shifted
] {sqed_outputs/sqed_scan_saddle.csv};
\end{axis}
\end{tikzpicture}%
}
\caption{Stationary shifted action $\left.S_{\rm eff}/N\right|_{\rm branch}$.}
\label{fig:app-action-heatmap}
\end{subfigure}
\hfill
\begin{subfigure}[t]{0.49\textwidth}
\centering
\resizebox{\linewidth}{!}{%
\begin{tikzpicture}
\begin{axis}[
    width=7cm,
    height=6cm,
    xlabel={$\lambda_x$},
    ylabel={$\lambda_y$},
    xmin=0, xmax=2,
    ymin=0, ymax=2,
    colorbar,
    colorbar style={ylabel={$F_*$}},
    colormap/viridis,
]
\addplot[
    scatter,
    only marks,
    mark=square*,
    mark size=2.2pt,
    scatter src=explicit,
] table[
    col sep=comma,
    x=alpha_x,
    y=alpha_y,
    meta=F
] {sqed_outputs/sqed_scan_saddle.csv};
\end{axis}
\end{tikzpicture}%
}
\caption{Saddle value $F_*$. The diagonal $\lambda_x=\lambda_y$ gives $F_*=0$, consistent with \eqref{eq:app-equal-F-eq}.}
\label{fig:app-f-heatmap}
\end{subfigure}
\caption{Heat maps over the $(\lambda_x,\lambda_y)$ plane: (\subref{fig:app-action-heatmap}) the stationary shifted action $\left.S_{\rm eff}/N\right|_{\rm branch}$ of \eqref{eq:app-equal-reduced}, and (\subref{fig:app-f-heatmap}) the corresponding saddle value $F_*$.}
\label{fig:app-heatmaps}
\end{figure}

Equation~\eqref{eq:app-equal-F-eq} is solved numerically by bisection, with the sums in \eqref{eq:app-equal-det-sum} and \eqref{eq:app-equal-F-eq} truncated at $l_{\max}=4000$ and $F$ discretized on a grid of $4001$ points in $[-\tfrac14,\tfrac14]$. The $(\lambda_x,\lambda_y)$ plane is sampled on a $41\times 41$ grid with $\lambda_{x,y}\in[0,2]$.
Figures \ref{fig:app-action-heatmap} and \ref{fig:app-f-heatmap} show, respectively, the stationary value of the shifted action \eqref{eq:app-equal-reduced} and the corresponding $F_*$ as functions of $(\lambda_x,\lambda_y)$.

\end{appendix}

\bibliographystyle{JHEP}
\bibliography{refs}

\end{document}